\documentclass[%
onecolumn,
superscriptaddress,
preprintnumbers,
nofootinbib,
notitlepage,
amsmath,amssymb,
aps,
prd,
floatfix,
longbibliography,
]{revtex4-1}

\usepackage{graphicx}
\usepackage{dcolumn}
\usepackage{bm}

\usepackage[dvipsnames]{xcolor}
\usepackage{hyperref}
\hypersetup{
    colorlinks=true,     
    linkcolor=blue,      
    citecolor=blue,      
    filecolor=blue,      
    urlcolor=blue        
}

\usepackage{mathtools} 
\usepackage{xspace} 
\usepackage{enumitem} 
\usepackage{braket}

\begin{document}

\title{Applications of Machine Learning to Lattice Quantum Field Theory\\Snowmass 2022 White Paper}

\newcommand{\getMITAffiliation}{\affiliation{Center for Theoretical Physics, Massachusetts Institute of Technology, Cambridge, MA 02139, USA}}
\newcommand{\getIAIFIAffiliation}{\affiliation{The NSF AI Institute for Artificial Intelligence and Fundamental Interactions}}
\newcommand{\getCQAAffiliation}{\affiliation{Co-Design Center for Quantum Advantage (C$\,^2$QA)}}
\newcommand{\getALCFAffiliation}{\affiliation{Leadership Computing Facility, Argonne National Laboratory, Argonne, IL 60439, USA}}
\newcommand{\getANLCPSAffiliation}{\affiliation{Computational Science Division, Argonne National Laboratory, Argonne, IL 60439, USA}}
\newcommand{\getSwanMathsAffiliation}{\affiliation{Department of Mathematics, Swansea University, Bay Campus, Swansea SA1 8EN, UK}}
\newcommand{\getSwanSAACAffiliation}{\affiliation{Swansea Academy of Advanced Computing, Swansea University, Bay Campus, Swansea SA1 8EN, UK}}
\newcommand{\getSwanPhysicsAffiliation}{\affiliation{Department of Physics, Swansea University, Swansea SA2 8PP, UK}}
\newcommand{\getECTAffiliation}{\affiliation{European Centre for Theoretical Studies in Nuclear Physics and Related Areas (ECT*) \& Fondazione Bruno Kessler Strada delle Tabarelle 286, 38123 Villazzano (TN), Italy}}

\author{Denis~Boyda}
\getALCFAffiliation
\getIAIFIAffiliation

\author{Salvatore Cal\`i}
\getMITAffiliation
\getIAIFIAffiliation

\author{Sam Foreman}
\getALCFAffiliation

\author{Lena Funcke}
\getMITAffiliation
\getIAIFIAffiliation
\getCQAAffiliation

\author{Daniel C. Hackett}
\thanks{Editor}
\email{dhackett@mit.edu}
\getMITAffiliation
\getIAIFIAffiliation

\author{Yin~Lin}
\getMITAffiliation
\getIAIFIAffiliation

\author{Gert~Aarts}
\getSwanPhysicsAffiliation
\getECTAffiliation

\author{Andrei Alexandru}
\affiliation{Physics Department, The George Washington University, Washington, DC 20052, USA}
\affiliation{Department of Physics, University of Maryland, College Park, MD 20742, USA}

\author{Xiao-Yong Jin}
\getALCFAffiliation
\getANLCPSAffiliation

\author{Biagio Lucini}
\getSwanMathsAffiliation
\getSwanSAACAffiliation

\author{Phiala E. Shanahan}
\getMITAffiliation
\getIAIFIAffiliation
\getCQAAffiliation

\preprint{MIT-CTP/5405}

\date{\today}

\begin{abstract}
There is great potential to apply machine learning in the area of numerical lattice quantum field theory, but full exploitation of that potential will require new strategies.
In this white paper for the Snowmass community planning process, we discuss the unique requirements of machine learning for lattice quantum field theory research and outline what is needed to enable exploration and deployment of this approach in the future.
\end{abstract}

\maketitle

\section{Introduction}

Lattice quantum field theory (LQFT), i.e., QFT discretized on a Euclidean spacetime lattice, is the only framework presently available to perform \textit{ab-initio} QFT calculations with fully controlled, systematically improvable uncertainties when the system of interest exhibits nonperturbative dynamics. Most importantly, this includes quantum chromodynamics (QCD), the component of the Standard Model governing the strong force (see e.g.~the recent topical issue of EPJA for a review~\cite{Brower:2019oor,Lehner:2019wvv,Kronfeld:2019nfb,Cirigliano:2019jig,Detmold:2019ghl,Bazavov:2019lgz,Joo:2019byq}). This important tool of modern physics continues to have great success in providing theory inputs necessary to understand and interpret experimental results. Notably, in quark flavor physics, Lattice QCD (LQCD) provides the hadronic matrix elements necessary to extract the Cabibbo–Kobayashi–Maskawa (CKM) parameters from experimental measurements and hence is critical to tests of CKM unitarity~\cite{Aoki:2021kgd}; in precision Higgs physics, LQCD provides the most precise determinations of the strong coupling constant and quark masses that enter in predictions of the branching ratios of the dominant decay modes of the Higgs boson~\cite{Aoki:2021kgd}; and lattice studies of QCD thermodynamics are essential to  interpret results from relativistic heavy-ion collisions~\cite{Ratti:2018ksb}.  

However, further progress is required to extend these successes of the LQCD approach to meet the needs of ongoing and near-future experimental efforts in high-energy (HEP) and nuclear physics. In particular, while the field is advancing rapidly, for many key applications the necessary LQCD calculations are limited by available computing power; with improvement in algorithms for LQCD, many more important contributions are on the horizon. For example, high-statistics, high-precision LQCD calculations of the hadronic vacuum polarization and hadronic light-by-light scattering contributions to $g-2$ of the muon could be critical to resolve tension between Standard Model predictions and experiments~\cite{Muong-2:2021ojo}. At the necessary (sub-percent) level of precision, isospin breaking and QED effects become important, and calculations in the theory with non-degenerate quark masses coupled to electromagnetism are affected by technical and conceptual problems. In addition, so-called disconnected quark diagrams contribute significantly at this level of precision. Stochastic methods have advanced sufficiently to provide access to the otherwise-intractable inverses of Dirac matrices involved in these contributions, but remain expensive and add large additional statistical uncertainties. Further advances in LQFT algorithms will contribute significantly to this pursuit~\cite{2021:gm2review}.

Similarly, LQCD has the potential to significantly impact the interpretation of the results of long-baseline neutrino experiments by providing constraints on nucleon and (together with effective field theory) nuclear matrix elements involved in neutrino scattering cross sections with heavy nuclei such as $^{12}$C, water, and $^{40}$Ar. Precise cross-section determination from nuclear models is crucial to reduce the final uncertainties of neutrino parameters, but the precision of these models depends on the uncertainties of their inputs. Many of the relevant matrix elements are difficult to constrain experimentally; in these cases, LQCD is the only viable approach to reduce their uncertainties and meet the needs of experiments \cite{Kronfeld:2019nfb,Meyer:2022mix,NuSTEC:2017hzk}. However, direct LQCD calculations of nuclei are computationally expensive and only systems with small atomic numbers are accessible at present. Calculations of larger nuclear systems require new developments to address increasingly severe signal-to-noise problems, which in some cases arise due to numerical sign problems~\cite{Wagman:2016bam}, as well as the high combinatoric complexity in Wick contractions. Similarly, computational cost is a significant factor in the computation of nucleon and nuclear matrix elements needed for the interpretation of dark matter direct detection~\cite{Cirigliano:2019jig} and neutrinoless double beta decay experiments~\cite{Engel:2016xgb,Dolinski:2019nrj}, as well as high-precision LQCD results for $g_A$ needed to resolve the outstanding discrepancy between results obtained from different experimental approaches to measuring the neutron lifetime~\cite{Gonzalez-Alonso:2018omy}. Extending the study of QCD at finite temperature to nonzero baryon density, as required for nuclear matter, neutron stars, and other phases of dense QCD also leads to a numerical sign problem, due to the complex nature of the Boltzmann weight in the grand-canonical formulation \cite{deForcrand:2009zkb,Aarts:2015tyj}.

More broadly, the past decades have seen applications of LQCD across all aspects of hadronic physics. These efforts have already yielded important insights, but in many cases further developments in LQFT technology are necessary to deliver results with the greatest possible impact. For example, the recent development of the quasi- and pseudo-PDF formalisms \cite{Ji:2013dva,Radyushkin:2017cyf} has enabled LQCD calculations of parton distribution functions (PDFs) and their generalizations, but these methods are still limited by available precision and lattice sizes. Alternately, PDFs may in principle be reconstructed from their moments, which can be computed directly on the lattice; however, issues with power divergent mixings demand presently impractical levels of statistical precision (or novel approaches \cite{Chambers:2017dov,Davoudi:2012ya}) for all but the lowest moments. Both of these approaches also involve ill-posed inverse problems, which also appear at finite temperature, in the reconstruction of spectral functions \cite{Asakawa:2000tr}, and the extraction of transport coefficients from numerically determined LQCD correlators \cite{Aarts:2002cc,Meyer:2011gj}.  Similarly, the correlation functions relevant to computing masses and matrix elements of higher excitations, such as resonances, glueballs, and exotic hadrons suffer from an overwhelming level of numerical noise at the sample sizes accessible with present methods. In some cases, as with nuclear correlation functions, this noise may arise due to a sign problem. Similar concerns apply for scattering amplitudes, which may be reconstructed from the spectra of correlation functions using finite-volume formalisms~\cite{Luscher:2009eq}. Besides QCD, LQFT methods have also been used for direct investigations of strongly coupled models for physics beyond the Standard Model \cite{degrandLatticeTestsStandard2016,Drach:2020qpj,Schaich:2018mmv}, such as supersymmetric gauge theories and models where the Higgs boson and/or dark matter are bound states of a new, as-yet unobserved confining force. Models with dilatonic Higgs bosons are particularly computationally demanding. LQFT methods have similarly been applied to address foundational questions in QFT \cite{Hernandez:2020tbc,Lucini:2012gg,Greensite:2003bk,Ripka:2003vv} and to systems of interest outside of particle physics \cite{Mathur:2016cko}. To summarize, in many of the cases described above, new physics results can be obtained by more efficient algorithms enabling higher statistical precision, while in others, it is clear that new ideas and novel approaches will be necessary. 

From its inception, advancements in LQFT have been achieved by combining increasing computing resources with formal and algorithmic developments. Such developments make new physics targets accessible to the LQFT framework not only by allowing more effective use of existing resources, but also by extending the formalism itself when known approaches are not effective or ill-defined.
In this context, emerging machine learning (ML) techniques offer an unprecedented new avenue for exploration, both in terms of increasing efficiency and for new, innovative formulations.

The past few years has seen promising exploratory applications of ML methods to all aspects of LQFT calculations, with many more in progress. To organize these efforts and the discussion below, we divide the workflow of an LQFT calculation into three sequential stages. For concreteness, these stages may be defined by the data types operated upon.
\begin{enumerate}[itemsep=4pt,parsep=0pt,topsep=0pt,leftmargin=*]
    \item \emph{Configuration generation} -- Samples of the lattice field degrees of freedom (field configurations) are drawn from the Boltzmann distribution defined by the lattice action. ML applications thus far include novel and/or improved sampling algorithms~\cite{Wang2017,Huang:2017,song2017nice,LiWang2018NNRG,levy2018generalizing,Albergo:2019eim,Wu:2019,Nicoli:2020evf,Nicoli:2020njz,Pawlowski:2018qxs,Kanwar:2020xzo,Boyda:2020hsi,Albergo:2021bna,Hackett:2021idh,Gabrie:2021tlu,DelDebbio:2021qwf,Wu:2021tfb,Foreman:2021ixr,Foreman:2021ljl,Foreman:2021rhs,Jin:2022bgq,Finkenrath:2022ogg,Bachtis:2020dmf,deHaan:2021erb,Cossu:2018pxj,Tanaka:2017niz,Tomiya:2021ywc,Bachtis:2021eww} and path-integral contour deformations for finite-density~\cite{Alexandru:2017czx,Lawrence:2021izu,Wynen:2020uzx}.
    \item \emph{Observable measurement} -- Quantities like correlation functions are evaluated over ensembles of field configurations. ML applications thus far include novel methods to extract thermodynamic observables~\cite{Nicoli:2020njz}, action parameter regression~\cite{Shanahan:2018vcv}, observable approximation~\cite{Favoni:2020reg,Bulusu:2021rqz,Matsumoto:2019jia}, design of new observables \cite{Boyda:2020nfh,Bachtis:2020fly,Bachtis:2020dmf,Palermo:2021jrf,Tan:2021cgs,Li:2017xaz,Wetzel:2017ooo,Alexandrou:2019hgt,Blucher:2020mjt,Yau:2020emg,Zhou:2018ill}, and path-integral contour deformations for baryonic correlators~\cite{Detmold:2021ulb}.
    \item \emph{Analysis} -- Physically interpretable results are extracted from observable measurements. ML applications thus far include cross-observable regression \cite{Yoon:2018krb,Zhang:2019qiq}, action parameter regression~\cite{Shanahan:2018vcv,Hudspith:2021iqu}, and new methods for ill-posed inverse problems \cite{Kades:2019wtd,Offler:2019eij,Horak:2021syv,Chen:2021giw,Wang:2021cqw,Shi:2022yqw}.
\end{enumerate}
As discussed further in Sec.~\ref{sec:ml-vs-lqft}, each of these stages involves different hierarchies of computational scale, requiring different resources and optimizations to apply ML at each stage. Further, each stage has different requirements to maintain full control over uncertainties.

The computational and formal aspects of each of these stages imply specific requirements on applications of ML to LQFT. In the remainder of this white paper, we first detail key prospects for, and challenges of ML for LQFT. From these, we infer what technical work will be required and discuss what resources the community will need to deploy to enable successful application of ML to LQFT. We close with an outlook on ML for LQFT.

\section{Prospects, Challenges, and Requirements of ML for LQFT}
\label{sec:ml-vs-lqft}

\subsection{Prospects}

There are important similarities between standard LQFT and ML methods; hence LQFT is already well-positioned to make effective use of emerging ML technologies. We organize the discussion below by these common features, and emphasize how they provide an unusual opportunity for interdisciplinary collaboration and great potential for cross-cutting impact.
\begin{itemize}[leftmargin=0pt]

\item[]
\textbf{Mathematical toolkit.} Statistics and linear algebra underpin the standard toolkits of both LQFT and ML. As with particle and condensed matter theory, once a common language is established, researchers from LQFT and those working on other applications of ML find they are concerned with related problems (see e.g.~the discussion of symmetry below). By the same analogy, there are likely methods known to one community that can be straightforwardly adapted to solve problems in the other, cf.\ the development of the HMC algorithm~\cite{Duane:1987de} in LQCD and its subsequent adoption in other computational fields, or e.g.~using interacting fields as building blocks for neural networks \cite{Bachtis:2021xoh}. The potential for mutual benefit offers an unprecedented opportunity for collaboration between the LQCD community and both fields outside physics and industry.

\item[]
\textbf{Computing requirements.}
ML and traditional LQFT methods are computationally intensive. The dominant computational cost in both cases is numerical linear algebra, which is effectively parallelized by GPUs. At present, LQFT requires more tightly interconnected nodes than typical ML applications, but very large models needing fast communications are becoming increasingly common. And, while modern LQFT calculations require expensive double precision (FP64) calculations to avoid unacceptable levels of round-off error, mixed precision algorithms allow efficient LQFT use of hardware with specific optimizations for low-precision formats (e.g.~single (FP32), half (FP16), or ML-specific formats (BFLOAT16, TF32)). This approach is already used to accelerate LQCD calculations on GPUs~\cite{Clark:2009wm} and has been extended to GPU tensor cores~\cite{Tu:2021dvv}; work is ongoing to exploit emerging ML-specialized hardware (e.g.~TPUs, IPUs). Thus, hardware that is well-suited for LQCD is likely to be well-suited for ML (and vice versa, increasingly), providing obvious benefits for ease of adoption of ML in LQFT. Given these similarities, ML for LQFT research may lead to faster and better deployment of ML-specialized hardware in traditional LQFT calculations, and computational science generically.

\item[] \textbf{Symmetry.} In an ML context, symmetries amount to constraints on a problem which must be learned unless they are incorporated explicitly. Symmetries and invariance/equivariance thus provide a generically useful framework for encoding a priori knowledge about a problem into an ML architecture, which often provides substantially improved performance. Architectures encoding LQFT-relevant symmetries have already been developed in various other contexts. For example, the success of convolutional neural networks in image processing has been substantially driven by their characteristic feature of encoding translational symmetry, which is also a property of LQFTs. Symmetric architectures developed for LQFT will likely be useful for other applications, opening up exciting new possibilities for physics-inspired ML architectures, with a wide range of applications both in academia and industry. For example, rotationally symmetric ML architectures could be useful for applications including 3D modeling, computer vision, and autonomous driving. Symmetric architectures are discussed in more generality in another white paper~\cite{in_prep_symm_wp}.

\item[]
\textbf{Software as a community resource.}
Significant effort in both the ML and LQFT communities has been dedicated to developing highly optimized open-source codebases.
In particular, under collaborative efforts the lattice gauge theory community has created several programming packages (e.g.~Grid~\cite{Boyle:2015tjk}, openQCD~\cite{openqcd}, QUDA~\cite{Clark:2009wm}, and the SciDAC stack~\cite{scidac}) that enable LQCD calculations to achieve high efficiency on state-of-the-art supercomputers. 
Similarly, the ML community has developed several independent frameworks, most popularly \texttt{PyTorch}~\cite{NEURIPS2019_9015}, \texttt{TensorFlow}~\cite{tensorflow2015-whitepaper}, and \texttt{Jax}~\cite{jax2018github}, which allow users to easily utilize GPUs. Another set of frameworks (including \texttt{horovod} \cite{sergeev2018horovod}, \texttt{DeepSpeed}~\cite{deepSpeed}, \texttt{PyTorch DDP} \cite{li2020pytorch}) were developed for scaling ML applications to multi-node machines.\footnote{These techniques rely on efficient communication between multiple workers, but have already demonstrated high performance sustaining more than one exaFLOPS in FP16 precision \cite{laanait2019exascale}.
However, they may be inefficient for some types of models and can be optimized much further, as in e.g.~the DeepSpeed~\cite{deepSpeed} library.} 
All of these frameworks are freely available and open-source, and thus may be combined to apply ML to LQFT without incurring the development cost of fully reimplementing either toolkit.
Integration of ML and LQFT software may also benefit traditional LQFT calculations. In particular, ML software frameworks are maintained by commercial vendors across a wide range of HPC hardware platforms, which could be leveraged to help address the problem of efficient portability of LQFT codes.
Likewise, an increased interest in ML libraries by the LQFT community can further accelerate their development.

\end{itemize}

\subsection{Challenges}

Despite these useful commonalities, there are a number of differences that present challenges for applying ML methods to LQFT problems. By considering them, we can assess what work will be required to bring ML into the standard LQFT toolkit.
\begin{itemize}[leftmargin=0pt]

\item[]
\textbf{Full control over uncertainties.} In many ML applications, formal guarantees of exactness are unnecessary and high-quality but approximate solutions are sufficient. Errors can be studied empirically on validation datasets, but characterization of systematic uncertainties may be difficult due to the black-box nature of neural networks.\footnote{Some recent activities in ML focus on out-of-distribution detection.} In contrast, LQFT applications demand correctness, i.e., that all sources of statistical and systematic uncertainty can be estimated reliably. This requirement manifests differently in each stage of the LQFT workflow:
\begin{enumerate}
    \item \textit{Configuration generation} requires provably exact sampling from the Boltzmann distribution defined by the lattice action, or at least sampling which provides sufficient statistical information to fully correct for violations of exactness. This imposes strict constraints on what ML approaches and architectures are applicable.
    \item \textit{Observable measurement} requires control over any violations of asymptotic unbiasedness induced by computing approximations of known observables (e.g.~by statistical estimation of a bias correction). Model dependence on training data may present additional concerns for asymptotics. Novel machine-learned observables must be carefully characterized and interpreted with caution.
    \item \textit{Analysis} more closely resembles typical ML applications. Most uses of ML at this stage will introduce model dependence, which must be treated as a source of systematic uncertainty and controlled for. This includes dependence on training data, which may induce important model-data correlations. Models with probabilistic interpretations may play a useful role. Estimates of uncertainty are often phrased in a Bayesian context, used in LQFT in e.g.\ spectral function reconstruction. Many concerns relevant for LQFT applications are discussed in more detail in another white paper on uncertainty quantification~\cite{in_prep_uncertainty_wp}.
\end{enumerate}

\item[]
\textbf{Differentiability.} The modern ML toolkit is designed around optimization with stochastic gradient descent, which necessitates that \emph{all} operations be automatically differentiable to enable backpropagation. This is not a typical requirement of LQFT calculations, which more often implement the necessary derivatives of the action manually to maximize performance, and so is not incorporated in any of the major LQFT codebases. Especially for applications to configuration generation and observable measurements, it may be necessary to build automatic differentiation into existing LQFT software, or reimplement parts of the LQFT toolkit inside ML frameworks.

\item[]
\textbf{Data hierarchies and computing models.} Typical ML applications operate on large volumes of data points, each of size $\sim$ KBs. Although this hierarchy applies in the analysis stage of LQFT calculations, it is very dissimilar to the relevant scales in state-of-the-art LQCD configuration generation and observable measurements, which involve processing relatively small volumes of field configurations and propagators of size $\sim$ GBs -- 100s of GBs. These large data sizes mean applying ML methods at state-of-the-art scale will require a higher degree of model parallelism than typical for ML, where data parallelism alone is often sufficient, and where the need for model parallelism is more often due to large model sizes.\footnote{In a data parallel scheme, parallel workers each apply the entire ML model to multiple pieces of data. This scheme involves relatively little communication between workers, requiring only the synchronization of gradients of model parameters once per backwards pass. In a model parallel scheme, the work of applying an ML model to a single piece of data is divided between multiple workers. This may be accomplished e.g.~by dividing the problem across sublattices, although other schemes like pipeline parallelism are possible.} Although the recent successes of very large ML models has led to software support for model parallelism, this infrastructure is still in its infancy. Engineering will be required to adapt ML software to make efficient use of high-performance computing resources. Further, problem-specific optimizations are more important for model parallel schemes, which clashes with the problem-agnostic approach employed by ML software frameworks. Applying ML to LQFT may require specialized software incorporating LQFT-specific optimizations.

\end{itemize}

\subsection{Requirements}

From these concerns, we conclude that successful application of ML techniques in lattice QCD will require significant efforts in two primary directions:
\begin{enumerate}[leftmargin=*]
    \item \textit{Exploratory research} -- The particular requirements of LQFT mean that out-of-the-box ML solutions often are not directly applicable. Applying ML to LQFT while retaining full control over uncertainties will require research and development of novel constructions with appropriate properties. Given the breadth of ML methods already available, and with the rapid ongoing development of ML, this amounts to a need for substantial exploratory research. The work required is experimental and inherently computationally demanding, especially when testing scalability on large systems.
    
    \item \textit{Software engineering} -- Following precedent, using ML-based methods at state-of-the-art scale will require development and long-term maintenance of publicly available software infrastructure by the LQFT community. This software must be portable and highly optimized for efficient use of available hardware resources, as well as well-tested, well-documented, and open source to ensure scientific validity and verifiability. Development can be accelerated by leveraging the large amount of work and domain expertise incorporated into existing LQFT and ML software frameworks. However, at present, these pieces of infrastructure are entirely separate, and as discussed above have been designed with different considerations (cf.~differentiability, data hierarchies and computing models). Delivering high performance in ML for LQFT at scale thus represents a substantial software engineering task, which will require experts with extensive knowledge in both domains.

\end{enumerate}

\section{Strategies to Enable ML for LQFT}

\begin{itemize}[leftmargin=0pt]
\item[] \textbf{Computing strategies.}  
    New policies for allocating computing time are needed to support exploratory algorithm development and applications of ML-based approaches. Usually, allocations are requested for a fixed amount of computing time, to run a specific set of computations on a predetermined schedule, and are granted based on what physics results the proposed computations are projected to enable. This does not match with the nature of exploratory ML research, which involves an iterative experimental process with each experiment guiding what computations are run next. Iteration timescales are typically of order days or weeks, much less than year-long computing allocations. More generally, even given a production-ready approach, present allocation policies are incompatible with any computation that involves training a model: until the training is actually carried out, the precise cost and outcome will not be known to high accuracy. Employing all talent available, and especially avoiding unfairly shutting out researchers from smaller and less-funded institutions, will require making computing resources broadly available.

\item[] \textbf{Community standards and resources.}
    As discussed above, ML for LQFT will require development and maintenance of specialized software toolkits. Just as with software, trained models should be treated as a community resource, particularly for at-scale applications where training may be expensive. Best practices must be established for what information is required to constitute a verifiable result~\cite{wilkinson2016fair}, likely involving sharing of code and models alongside publications. To these ends, centralized infrastructure will be needed to enable sharing of trained models and the code to use them.

\item[] \textbf{Career paths.} 
    Permanent positions in ML for physics must be made available to support the exploratory research and software engineering described above. This includes both traditional academic jobs and positions for specialists in technical roles, such as Research Software Engineers (RSEs). Increased support for research scientists working on ML for physics will promote building a vibrant interdisciplinary community and bridge gaps between different subfields of physics and beyond. Given the utility of ML outside academia, these concerns are especially important to retain talented early-career researchers, particularly those engaged in valuable but highly technical work. Policies and practices should be adjusted for a definition of physics research and education that incorporates computer science and applied mathematics relevant to ML. Cross-disciplinary collaborations may provide rapid progress as well as access to substantial non-traditional funding and computational resources, including from industrial partners. Hiring and graduate admissions should consider applicants from outside physics; it may be that the best-prepared candidate for ML-based research is a student with an undergraduate degree in computer science, or a researcher from another field or even industry. This applies especially in education, as discussed in greater detail in another white paper~\cite{in_prep_education_wp}. While specialized degrees can play a useful role, ML for physics is a legitimate topic for a physics degree. Qualifying exam practices should allow for students specialized in computation. Physics departments should incorporate computing and ML in their curricula; besides supporting the development of ML for the physics community, this will benefit the majority of students who go on to careers outside research.

\end{itemize}

\section{Outlook}

From vector machines to BlueGene to GPUs, the LQFT community has a long history of leadership in rapidly adopting new computing technologies, and driving their development as they emerge. The community now has the opportunity to position itself for similar leadership in ML. Promising proof-of-principle results across every aspect of the LQFT workflow and rapidly growing engagement in work at this intersection both illustrate potential to deliver transformative advances in the immediate future.

Realizing this potential will require intentional investment of human and computing resources by the community in ways distinct from those that have driven traditional algorithms research. Bringing ML to bear will require extensive, computationally demanding exploratory research incorporating developments from the broader field of ML, requiring updates to the way computing resources are allocated. When promising methods are identified, deploying them at state-of-the-art scale will require software engineering to meet the existing high standards for performance.

The development and deployment of novel ML algorithms for LQFT has great potential for cross-cutting impact. Besides its widespread adoption in industry, ML methods have already been applied to various problems in particle/nuclear physics~\cite{Feickert:2021ajf,Boehnlein:2021eym,tanaka2021deep},  astronomy~\cite{Cuoco:2020ogp,baron2019machine}, condensed-matter physics~\cite{Bedolla-Montiel:2020rio,Bedolla-Montiel:2020rio},  computational fluid dynamics~\cite{doi:10.1146/annurev-fluid-010719-060214,Kochkove2101784118}, quantum chemistry~\cite{schutt2020machine}, and many further fields of physics and beyond. As outlined in this white paper, applications to LQFT bring specific challenges to the forefront, such as symmetries and the need for provably exact algorithms. The solutions to these challenges may provide opportunities not only for science but industrial applications as well. Moreover, most of these considerations would be equally beneficial for other emerging computational research directions, including tensor networks and quantum computing.

Developing the interdisciplinary workforce who will carry out this work will require investment and advocacy by the LQFT community. For example, early career researchers are responsible for much of the ongoing research in ML for LQFT; the community risks losing this talent pool if it cannot provide permanent positions to retain them, either by opening physics positions to researchers with such an algorithmic focus, by opening computer science positions to researchers with a focus on physics problems, or by creating permanent positions at the interdisciplinary boundaries. This talent pool can be further expanded through the inclusion of computing and ML in physics curricula, and more generally opening the field of physics to admit the specialization necessary for ML-based research. These same concerns apply not only for LQFT, but for ML for physics in general. Nevertheless, if these challenges can be successfully negotiated, the interdisciplinary nature and generalizability of ML methods provide a unique opportunity for collaboration between fields and with industry, opening the doors to new intellectual, computational, and funding resources which may have significant impact on the state-of-the-art in HEP theory.

\begin{acknowledgements}
DB, LF, DCH, YL, SC, and PES are supported in part by the U.S.\ Department of Energy, Office of Science, Office of Nuclear Physics, under grant Contract Number DE-SC0011090. PES is additionally supported by the National Science Foundation under EAGER grant 2035015, by the U.S.\ DOE Early Career Award DE-SC0021006, by a NEC research award, and by the Carl G and Shirley Sontheimer Research Fund. YL is also supported by the National Science Foundation award PHY-2019786. LF is also supported by the U.S.\ Department of Energy, Office of Science, National Quantum Information Science Research Centers, Co-design Center for Quantum Advantage (C$^2$QA) under contract number DE-SC0012704, by the DOE QuantiSED Consortium under subcontract number 675352, by the National Science Foundation under Cooperative Agreement PHY-2019786 (The NSF AI Institute for Artificial Intelligence and Fundamental Interactions, http://iaifi.org/), and by the U.S.\ Department of Energy, Office of Science, Office of Nuclear Physics under grant contract number DE-SC0021006. DB, SF and XJ were supported by the Argonne Leadership Computing Facility, which is a U.S. Department of Energy Office of Science User Facility operated under contract DE-AC02-06CH11357. SF and XJ were additionally supported by the Exascale Computing Project (17-SC-20-SC), a collaborative effort of the U.S. Department of Energy Office of Science and the National Nuclear Security Administration. GA and BL are supported in part by the UKRI Science and Technology Facilities Council (STFC) Consolidated Grant ST/T000813/1. The work of BL is further supported in part by the Royal Society Wolfson Research Merit Award WM170010 and by the Leverhulme Foundation Research Fellowship RF-2020-461/9. 
AA is supported in part by U.S. DOE Grant No. DE-FG02-95ER40907.
\end{acknowledgements}

\bibliography{main}

\begin{thebibliography}{117}%
\makeatletter
\providecommand \@ifxundefined [1]{%
 \@ifx{#1\undefined}
}%
\providecommand \@ifnum [1]{%
 \ifnum #1\expandafter \@firstoftwo
 \else \expandafter \@secondoftwo
 \fi
}%
\providecommand \@ifx [1]{%
 \ifx #1\expandafter \@firstoftwo
 \else \expandafter \@secondoftwo
 \fi
}%
\providecommand \natexlab [1]{#1}%
\providecommand \enquote  [1]{``#1''}%
\providecommand \bibnamefont  [1]{#1}%
\providecommand \bibfnamefont [1]{#1}%
\providecommand \citenamefont [1]{#1}%
\providecommand \href@noop [0]{\@secondoftwo}%
\providecommand \href [0]{\begingroup \@sanitize@url \@href}%
\providecommand \@href[1]{\@@startlink{#1}\@@href}%
\providecommand \@@href[1]{\endgroup#1\@@endlink}%
\providecommand \@sanitize@url [0]{\catcode `\\12\catcode `\$12\catcode
  `\&12\catcode `\#12\catcode `\^12\catcode `\_12\catcode `\%12\relax}%
\providecommand \@@startlink[1]{}%
\providecommand \@@endlink[0]{}%
\providecommand \url  [0]{\begingroup\@sanitize@url \@url }%
\providecommand \@url [1]{\endgroup\@href {#1}{\urlprefix }}%
\providecommand \urlprefix  [0]{URL }%
\providecommand \Eprint [0]{\href }%
\providecommand \doibase [0]{http://dx.doi.org/}%
\providecommand \selectlanguage [0]{\@gobble}%
\providecommand \bibinfo  [0]{\@secondoftwo}%
\providecommand \bibfield  [0]{\@secondoftwo}%
\providecommand \translation [1]{[#1]}%
\providecommand \BibitemOpen [0]{}%
\providecommand \bibitemStop [0]{}%
\providecommand \bibitemNoStop [0]{.\EOS\space}%
\providecommand \EOS [0]{\spacefactor3000\relax}%
\providecommand \BibitemShut  [1]{\csname bibitem#1\endcsname}%
\let\auto@bib@innerbib\@empty
\bibitem [{\citenamefont {Brower}\ \emph {et~al.}(2019)\citenamefont {Brower},
  \citenamefont {Hasenfratz}, \citenamefont {Neil}, \citenamefont {Catterall},
  \citenamefont {Fleming}, \citenamefont {Giedt}, \citenamefont {Rinaldi},
  \citenamefont {Schaich}, \citenamefont {Weinberg},\ and\ \citenamefont
  {Witzel}}]{Brower:2019oor}%
  \BibitemOpen
  \bibfield  {author} {\bibinfo {author} {\bibfnamefont {Richard~C.}\
  \bibnamefont {Brower}}, \bibinfo {author} {\bibfnamefont {Anna}\ \bibnamefont
  {Hasenfratz}}, \bibinfo {author} {\bibfnamefont {Ethan~T.}\ \bibnamefont
  {Neil}}, \bibinfo {author} {\bibfnamefont {Simon}\ \bibnamefont {Catterall}},
  \bibinfo {author} {\bibfnamefont {George}\ \bibnamefont {Fleming}}, \bibinfo
  {author} {\bibfnamefont {Joel}\ \bibnamefont {Giedt}}, \bibinfo {author}
  {\bibfnamefont {Enrico}\ \bibnamefont {Rinaldi}}, \bibinfo {author}
  {\bibfnamefont {David}\ \bibnamefont {Schaich}}, \bibinfo {author}
  {\bibfnamefont {Evan}\ \bibnamefont {Weinberg}}, \ and\ \bibinfo {author}
  {\bibfnamefont {Oliver}\ \bibnamefont {Witzel}} (\bibinfo {collaboration}
  {USQCD}),\ }\bibfield  {title} {\enquote {\bibinfo {title} {{Lattice Gauge
  Theory for Physics Beyond the Standard Model}},}\ }\href {\doibase
  10.1140/epja/i2019-12901-5} {\bibfield  {journal} {\bibinfo  {journal} {Eur.
  Phys. J. A}\ }\textbf {\bibinfo {volume} {55}},\ \bibinfo {pages} {198}
  (\bibinfo {year} {2019})},\ \Eprint {http://arxiv.org/abs/1904.09964}
  {arXiv:1904.09964 [hep-lat]} \BibitemShut {NoStop}%
\bibitem [{\citenamefont {Lehner}\ \emph {et~al.}(2019)\citenamefont {Lehner}
  \emph {et~al.}}]{Lehner:2019wvv}%
  \BibitemOpen
  \bibfield  {author} {\bibinfo {author} {\bibfnamefont {Christoph}\
  \bibnamefont {Lehner}} \emph {et~al.} (\bibinfo {collaboration} {USQCD}),\
  }\bibfield  {title} {\enquote {\bibinfo {title} {{Opportunities for Lattice
  QCD in Quark and Lepton Flavor Physics}},}\ }\href {\doibase
  10.1140/epja/i2019-12891-2} {\bibfield  {journal} {\bibinfo  {journal} {Eur.
  Phys. J. A}\ }\textbf {\bibinfo {volume} {55}},\ \bibinfo {pages} {195}
  (\bibinfo {year} {2019})},\ \Eprint {http://arxiv.org/abs/1904.09479}
  {arXiv:1904.09479 [hep-lat]} \BibitemShut {NoStop}%
\bibitem [{\citenamefont {Kronfeld}\ \emph {et~al.}(2019)\citenamefont
  {Kronfeld}, \citenamefont {Richards}, \citenamefont {Detmold}, \citenamefont
  {Gupta}, \citenamefont {Lin}, \citenamefont {Liu}, \citenamefont {Meyer},
  \citenamefont {Sufian},\ and\ \citenamefont {Syritsyn}}]{Kronfeld:2019nfb}%
  \BibitemOpen
  \bibfield  {author} {\bibinfo {author} {\bibfnamefont {Andreas~S.}\
  \bibnamefont {Kronfeld}}, \bibinfo {author} {\bibfnamefont {David~G.}\
  \bibnamefont {Richards}}, \bibinfo {author} {\bibfnamefont {William}\
  \bibnamefont {Detmold}}, \bibinfo {author} {\bibfnamefont {Rajan}\
  \bibnamefont {Gupta}}, \bibinfo {author} {\bibfnamefont {Huey-Wen}\
  \bibnamefont {Lin}}, \bibinfo {author} {\bibfnamefont {Keh-Fei}\ \bibnamefont
  {Liu}}, \bibinfo {author} {\bibfnamefont {Aaron~S.}\ \bibnamefont {Meyer}},
  \bibinfo {author} {\bibfnamefont {Raza}\ \bibnamefont {Sufian}}, \ and\
  \bibinfo {author} {\bibfnamefont {Sergey}\ \bibnamefont {Syritsyn}} (\bibinfo
  {collaboration} {USQCD}),\ }\bibfield  {title} {\enquote {\bibinfo {title}
  {{Lattice QCD and Neutrino-Nucleus Scattering}},}\ }\href {\doibase
  10.1140/epja/i2019-12916-x} {\bibfield  {journal} {\bibinfo  {journal} {Eur.
  Phys. J. A}\ }\textbf {\bibinfo {volume} {55}},\ \bibinfo {pages} {196}
  (\bibinfo {year} {2019})},\ \Eprint {http://arxiv.org/abs/1904.09931}
  {arXiv:1904.09931 [hep-lat]} \BibitemShut {NoStop}%
\bibitem [{\citenamefont {Cirigliano}\ \emph {et~al.}(2019)\citenamefont
  {Cirigliano}, \citenamefont {Davoudi}, \citenamefont {Bhattacharya},
  \citenamefont {Izubuchi}, \citenamefont {Shanahan}, \citenamefont
  {Syritsyn},\ and\ \citenamefont {Wagman}}]{Cirigliano:2019jig}%
  \BibitemOpen
  \bibfield  {author} {\bibinfo {author} {\bibfnamefont {Vincenzo}\
  \bibnamefont {Cirigliano}}, \bibinfo {author} {\bibfnamefont {Zohreh}\
  \bibnamefont {Davoudi}}, \bibinfo {author} {\bibfnamefont {Tanmoy}\
  \bibnamefont {Bhattacharya}}, \bibinfo {author} {\bibfnamefont {Taku}\
  \bibnamefont {Izubuchi}}, \bibinfo {author} {\bibfnamefont {Phiala~E.}\
  \bibnamefont {Shanahan}}, \bibinfo {author} {\bibfnamefont {Sergey}\
  \bibnamefont {Syritsyn}}, \ and\ \bibinfo {author} {\bibfnamefont
  {Michael~L.}\ \bibnamefont {Wagman}} (\bibinfo {collaboration} {USQCD}),\
  }\bibfield  {title} {\enquote {\bibinfo {title} {{The Role of Lattice QCD in
  Searches for Violations of Fundamental Symmetries and Signals for New
  Physics}},}\ }\href {\doibase 10.1140/epja/i2019-12889-8} {\bibfield
  {journal} {\bibinfo  {journal} {Eur. Phys. J. A}\ }\textbf {\bibinfo {volume}
  {55}},\ \bibinfo {pages} {197} (\bibinfo {year} {2019})},\ \Eprint
  {http://arxiv.org/abs/1904.09704} {arXiv:1904.09704 [hep-lat]} \BibitemShut
  {NoStop}%
\bibitem [{\citenamefont {Detmold}\ \emph {et~al.}(2019)\citenamefont
  {Detmold}, \citenamefont {Edwards}, \citenamefont {Dudek}, \citenamefont
  {Engelhardt}, \citenamefont {Lin}, \citenamefont {Meinel}, \citenamefont
  {Orginos},\ and\ \citenamefont {Shanahan}}]{Detmold:2019ghl}%
  \BibitemOpen
  \bibfield  {author} {\bibinfo {author} {\bibfnamefont {William}\ \bibnamefont
  {Detmold}}, \bibinfo {author} {\bibfnamefont {Robert~G.}\ \bibnamefont
  {Edwards}}, \bibinfo {author} {\bibfnamefont {Jozef~J.}\ \bibnamefont
  {Dudek}}, \bibinfo {author} {\bibfnamefont {Michael}\ \bibnamefont
  {Engelhardt}}, \bibinfo {author} {\bibfnamefont {Huey-Wen}\ \bibnamefont
  {Lin}}, \bibinfo {author} {\bibfnamefont {Stefan}\ \bibnamefont {Meinel}},
  \bibinfo {author} {\bibfnamefont {Kostas}\ \bibnamefont {Orginos}}, \ and\
  \bibinfo {author} {\bibfnamefont {Phiala}\ \bibnamefont {Shanahan}} (\bibinfo
  {collaboration} {USQCD}),\ }\bibfield  {title} {\enquote {\bibinfo {title}
  {{Hadrons and Nuclei}},}\ }\href {\doibase 10.1140/epja/i2019-12902-4}
  {\bibfield  {journal} {\bibinfo  {journal} {Eur. Phys. J. A}\ }\textbf
  {\bibinfo {volume} {55}},\ \bibinfo {pages} {193} (\bibinfo {year} {2019})},\
  \Eprint {http://arxiv.org/abs/1904.09512} {arXiv:1904.09512 [hep-lat]}
  \BibitemShut {NoStop}%
\bibitem [{\citenamefont {Bazavov}\ \emph {et~al.}(2019)\citenamefont
  {Bazavov}, \citenamefont {Karsch}, \citenamefont {Mukherjee},\ and\
  \citenamefont {Petreczky}}]{Bazavov:2019lgz}%
  \BibitemOpen
  \bibfield  {author} {\bibinfo {author} {\bibfnamefont {Alexei}\ \bibnamefont
  {Bazavov}}, \bibinfo {author} {\bibfnamefont {Frithjof}\ \bibnamefont
  {Karsch}}, \bibinfo {author} {\bibfnamefont {Swagato}\ \bibnamefont
  {Mukherjee}}, \ and\ \bibinfo {author} {\bibfnamefont {Peter}\ \bibnamefont
  {Petreczky}} (\bibinfo {collaboration} {USQCD}),\ }\bibfield  {title}
  {\enquote {\bibinfo {title} {{Hot-dense Lattice QCD: USQCD whitepaper
  2018}},}\ }\href {\doibase 10.1140/epja/i2019-12922-0} {\bibfield  {journal}
  {\bibinfo  {journal} {Eur. Phys. J. A}\ }\textbf {\bibinfo {volume} {55}},\
  \bibinfo {pages} {194} (\bibinfo {year} {2019})},\ \Eprint
  {http://arxiv.org/abs/1904.09951} {arXiv:1904.09951 [hep-lat]} \BibitemShut
  {NoStop}%
\bibitem [{\citenamefont {Jo\'o}\ \emph {et~al.}(2019)\citenamefont {Jo\'o},
  \citenamefont {Jung}, \citenamefont {Christ}, \citenamefont {Detmold},
  \citenamefont {Edwards}, \citenamefont {Savage},\ and\ \citenamefont
  {Shanahan}}]{Joo:2019byq}%
  \BibitemOpen
  \bibfield  {author} {\bibinfo {author} {\bibfnamefont {B\'alint}\
  \bibnamefont {Jo\'o}}, \bibinfo {author} {\bibfnamefont {Chulwoo}\
  \bibnamefont {Jung}}, \bibinfo {author} {\bibfnamefont {Norman~H.}\
  \bibnamefont {Christ}}, \bibinfo {author} {\bibfnamefont {William}\
  \bibnamefont {Detmold}}, \bibinfo {author} {\bibfnamefont {Robert}\
  \bibnamefont {Edwards}}, \bibinfo {author} {\bibfnamefont {Martin}\
  \bibnamefont {Savage}}, \ and\ \bibinfo {author} {\bibfnamefont {Phiala}\
  \bibnamefont {Shanahan}} (\bibinfo {collaboration} {USQCD}),\ }\bibfield
  {title} {\enquote {\bibinfo {title} {{Status and Future Perspectives for
  Lattice Gauge Theory Calculations to the Exascale and Beyond}},}\ }\href
  {\doibase 10.1140/epja/i2019-12919-7} {\bibfield  {journal} {\bibinfo
  {journal} {Eur. Phys. J. A}\ }\textbf {\bibinfo {volume} {55}},\ \bibinfo
  {pages} {199} (\bibinfo {year} {2019})},\ \Eprint
  {http://arxiv.org/abs/1904.09725} {arXiv:1904.09725 [hep-lat]} \BibitemShut
  {NoStop}%
\bibitem [{\citenamefont {Aoki}\ \emph {et~al.}(2021)\citenamefont {Aoki} \emph
  {et~al.}}]{Aoki:2021kgd}%
  \BibitemOpen
  \bibfield  {author} {\bibinfo {author} {\bibfnamefont {Y.}~\bibnamefont
  {Aoki}} \emph {et~al.},\ }\bibfield  {title} {\enquote {\bibinfo {title}
  {{FLAG Review 2021}},}\ }\href@noop {} {\  (\bibinfo {year} {2021})},\
  \Eprint {http://arxiv.org/abs/2111.09849} {arXiv:2111.09849 [hep-lat]}
  \BibitemShut {NoStop}%
\bibitem [{\citenamefont {Ratti}(2018)}]{Ratti:2018ksb}%
  \BibitemOpen
  \bibfield  {author} {\bibinfo {author} {\bibfnamefont {Claudia}\ \bibnamefont
  {Ratti}},\ }\bibfield  {title} {\enquote {\bibinfo {title} {{Lattice QCD and
  heavy ion collisions: a review of recent progress}},}\ }\href {\doibase
  10.1088/1361-6633/aabb97} {\bibfield  {journal} {\bibinfo  {journal} {Rept.
  Prog. Phys.}\ }\textbf {\bibinfo {volume} {81}},\ \bibinfo {pages} {084301}
  (\bibinfo {year} {2018})},\ \Eprint {http://arxiv.org/abs/1804.07810}
  {arXiv:1804.07810 [hep-lat]} \BibitemShut {NoStop}%
\bibitem [{\citenamefont {Abi}\ \emph {et~al.}(2021)\citenamefont {Abi} \emph
  {et~al.}}]{Muong-2:2021ojo}%
  \BibitemOpen
  \bibfield  {author} {\bibinfo {author} {\bibfnamefont {B.}~\bibnamefont
  {Abi}} \emph {et~al.} (\bibinfo {collaboration} {Muon g-2}),\ }\bibfield
  {title} {\enquote {\bibinfo {title} {{Measurement of the Positive Muon
  Anomalous Magnetic Moment to 0.46 ppm}},}\ }\href {\doibase
  10.1103/PhysRevLett.126.141801} {\bibfield  {journal} {\bibinfo  {journal}
  {Phys. Rev. Lett.}\ }\textbf {\bibinfo {volume} {126}},\ \bibinfo {pages}
  {141801} (\bibinfo {year} {2021})},\ \Eprint
  {http://arxiv.org/abs/2104.03281} {arXiv:2104.03281 [hep-ex]} \BibitemShut
  {NoStop}%
\bibitem [{\citenamefont {El-Khadra}(2021)}]{2021:gm2review}%
  \BibitemOpen
  \bibfield  {author} {\bibinfo {author} {\bibfnamefont {A.~X.}\ \bibnamefont
  {El-Khadra}},\ }\href
  {https://indico.cern.ch/event/1006302/contributions/4366872/attachments/2290248/3893802/El-Khadra_Lattice2021.pdf}
  {\enquote {\bibinfo {title} {Review of muon g-2},}\ } (\bibinfo {year}
  {2021}),\ \bibinfo {note} {{The 38th International Symposium on Lattice Field
  Theory}}\BibitemShut {NoStop}%
\bibitem [{\citenamefont {Meyer}\ \emph {et~al.}(2022)\citenamefont {Meyer},
  \citenamefont {Walker-Loud},\ and\ \citenamefont
  {Wilkinson}}]{Meyer:2022mix}%
  \BibitemOpen
  \bibfield  {author} {\bibinfo {author} {\bibfnamefont {Aaron~S.}\
  \bibnamefont {Meyer}}, \bibinfo {author} {\bibfnamefont {Andr\'e}\
  \bibnamefont {Walker-Loud}}, \ and\ \bibinfo {author} {\bibfnamefont
  {Callum}\ \bibnamefont {Wilkinson}},\ }\href@noop {} {\enquote {\bibinfo
  {title} {{Status of Lattice QCD Determination of Nucleon Form Factors and
  their Relevance for the Few-GeV Neutrino Program}},}\ } (\bibinfo {year}
  {2022}),\ \Eprint {http://arxiv.org/abs/2201.01839} {arXiv:2201.01839
  [hep-lat]} \BibitemShut {NoStop}%
\bibitem [{\citenamefont {Alvarez-Ruso}\ \emph {et~al.}(2018)\citenamefont
  {Alvarez-Ruso} \emph {et~al.}}]{NuSTEC:2017hzk}%
  \BibitemOpen
  \bibfield  {author} {\bibinfo {author} {\bibfnamefont {L.}~\bibnamefont
  {Alvarez-Ruso}} \emph {et~al.} (\bibinfo {collaboration} {NuSTEC}),\
  }\bibfield  {title} {\enquote {\bibinfo {title} {{NuSTEC White Paper: Status
  and challenges of neutrino\textendash{}nucleus scattering}},}\ }\href
  {\doibase 10.1016/j.ppnp.2018.01.006} {\bibfield  {journal} {\bibinfo
  {journal} {Prog. Part. Nucl. Phys.}\ }\textbf {\bibinfo {volume} {100}},\
  \bibinfo {pages} {1--68} (\bibinfo {year} {2018})},\ \Eprint
  {http://arxiv.org/abs/1706.03621} {arXiv:1706.03621 [hep-ph]} \BibitemShut
  {NoStop}%
\bibitem [{\citenamefont {Wagman}\ and\ \citenamefont
  {Savage}(2017)}]{Wagman:2016bam}%
  \BibitemOpen
  \bibfield  {author} {\bibinfo {author} {\bibfnamefont {Michael~L.}\
  \bibnamefont {Wagman}}\ and\ \bibinfo {author} {\bibfnamefont {Martin~J.}\
  \bibnamefont {Savage}},\ }\bibfield  {title} {\enquote {\bibinfo {title}
  {{Statistics of baryon correlation functions in lattice QCD}},}\ }\href
  {\doibase 10.1103/PhysRevD.96.114508} {\bibfield  {journal} {\bibinfo
  {journal} {Phys. Rev. D}\ }\textbf {\bibinfo {volume} {96}},\ \bibinfo
  {pages} {114508} (\bibinfo {year} {2017})},\ \Eprint
  {http://arxiv.org/abs/1611.07643} {arXiv:1611.07643 [hep-lat]} \BibitemShut
  {NoStop}%
\bibitem [{\citenamefont {Engel}\ and\ \citenamefont
  {Men\'endez}(2017)}]{Engel:2016xgb}%
  \BibitemOpen
  \bibfield  {author} {\bibinfo {author} {\bibfnamefont {Jonathan}\
  \bibnamefont {Engel}}\ and\ \bibinfo {author} {\bibfnamefont {Javier}\
  \bibnamefont {Men\'endez}},\ }\bibfield  {title} {\enquote {\bibinfo {title}
  {{Status and Future of Nuclear Matrix Elements for Neutrinoless Double-Beta
  Decay: A Review}},}\ }\href {\doibase 10.1088/1361-6633/aa5bc5} {\bibfield
  {journal} {\bibinfo  {journal} {Rept. Prog. Phys.}\ }\textbf {\bibinfo
  {volume} {80}},\ \bibinfo {pages} {046301} (\bibinfo {year} {2017})},\
  \Eprint {http://arxiv.org/abs/1610.06548} {arXiv:1610.06548 [nucl-th]}
  \BibitemShut {NoStop}%
\bibitem [{\citenamefont {Dolinski}\ \emph {et~al.}(2019)\citenamefont
  {Dolinski}, \citenamefont {Poon},\ and\ \citenamefont
  {Rodejohann}}]{Dolinski:2019nrj}%
  \BibitemOpen
  \bibfield  {author} {\bibinfo {author} {\bibfnamefont {Michelle~J.}\
  \bibnamefont {Dolinski}}, \bibinfo {author} {\bibfnamefont {Alan W.~P.}\
  \bibnamefont {Poon}}, \ and\ \bibinfo {author} {\bibfnamefont {Werner}\
  \bibnamefont {Rodejohann}},\ }\bibfield  {title} {\enquote {\bibinfo {title}
  {{Neutrinoless Double-Beta Decay: Status and Prospects}},}\ }\href {\doibase
  10.1146/annurev-nucl-101918-023407} {\bibfield  {journal} {\bibinfo
  {journal} {Ann. Rev. Nucl. Part. Sci.}\ }\textbf {\bibinfo {volume} {69}},\
  \bibinfo {pages} {219--251} (\bibinfo {year} {2019})},\ \Eprint
  {http://arxiv.org/abs/1902.04097} {arXiv:1902.04097 [nucl-ex]} \BibitemShut
  {NoStop}%
\bibitem [{\citenamefont {Gonz\'alez-Alonso}\ \emph {et~al.}(2019)\citenamefont
  {Gonz\'alez-Alonso}, \citenamefont {Naviliat-Cuncic},\ and\ \citenamefont
  {Severijns}}]{Gonzalez-Alonso:2018omy}%
  \BibitemOpen
  \bibfield  {author} {\bibinfo {author} {\bibfnamefont {Martin}\ \bibnamefont
  {Gonz\'alez-Alonso}}, \bibinfo {author} {\bibfnamefont {Oscar}\ \bibnamefont
  {Naviliat-Cuncic}}, \ and\ \bibinfo {author} {\bibfnamefont {Nathal}\
  \bibnamefont {Severijns}},\ }\bibfield  {title} {\enquote {\bibinfo {title}
  {{New physics searches in nuclear and neutron $\beta$ decay}},}\ }\href
  {\doibase 10.1016/j.ppnp.2018.08.002} {\bibfield  {journal} {\bibinfo
  {journal} {Prog. Part. Nucl. Phys.}\ }\textbf {\bibinfo {volume} {104}},\
  \bibinfo {pages} {165--223} (\bibinfo {year} {2019})},\ \Eprint
  {http://arxiv.org/abs/1803.08732} {arXiv:1803.08732 [hep-ph]} \BibitemShut
  {NoStop}%
\bibitem [{\citenamefont {de~Forcrand}(2009)}]{deForcrand:2009zkb}%
  \BibitemOpen
  \bibfield  {author} {\bibinfo {author} {\bibfnamefont {Philippe}\
  \bibnamefont {de~Forcrand}},\ }\bibfield  {title} {\enquote {\bibinfo {title}
  {{Simulating QCD at finite density}},}\ }\href {\doibase 10.22323/1.091.0010}
  {\bibfield  {journal} {\bibinfo  {journal} {PoS}\ }\textbf {\bibinfo {volume}
  {LAT2009}},\ \bibinfo {pages} {010} (\bibinfo {year} {2009})},\ \Eprint
  {http://arxiv.org/abs/1005.0539} {arXiv:1005.0539 [hep-lat]} \BibitemShut
  {NoStop}%
\bibitem [{\citenamefont {Aarts}(2016)}]{Aarts:2015tyj}%
  \BibitemOpen
  \bibfield  {author} {\bibinfo {author} {\bibfnamefont {Gert}\ \bibnamefont
  {Aarts}},\ }\bibfield  {title} {\enquote {\bibinfo {title} {{Introductory
  lectures on lattice QCD at nonzero baryon number}},}\ }\href {\doibase
  10.1088/1742-6596/706/2/022004} {\bibfield  {journal} {\bibinfo  {journal}
  {J. Phys. Conf. Ser.}\ }\textbf {\bibinfo {volume} {706}},\ \bibinfo {pages}
  {022004} (\bibinfo {year} {2016})},\ \Eprint
  {http://arxiv.org/abs/1512.05145} {arXiv:1512.05145 [hep-lat]} \BibitemShut
  {NoStop}%
\bibitem [{\citenamefont {Ji}(2013)}]{Ji:2013dva}%
  \BibitemOpen
  \bibfield  {author} {\bibinfo {author} {\bibfnamefont {Xiangdong}\
  \bibnamefont {Ji}},\ }\bibfield  {title} {\enquote {\bibinfo {title} {{Parton
  Physics on a Euclidean Lattice}},}\ }\href {\doibase
  10.1103/PhysRevLett.110.262002} {\bibfield  {journal} {\bibinfo  {journal}
  {Phys. Rev. Lett.}\ }\textbf {\bibinfo {volume} {110}},\ \bibinfo {pages}
  {262002} (\bibinfo {year} {2013})},\ \Eprint {http://arxiv.org/abs/1305.1539}
  {arXiv:1305.1539 [hep-ph]} \BibitemShut {NoStop}%
\bibitem [{\citenamefont {Radyushkin}(2017)}]{Radyushkin:2017cyf}%
  \BibitemOpen
  \bibfield  {author} {\bibinfo {author} {\bibfnamefont {A.~V.}\ \bibnamefont
  {Radyushkin}},\ }\bibfield  {title} {\enquote {\bibinfo {title}
  {{Quasi-parton distribution functions, momentum distributions, and
  pseudo-parton distribution functions}},}\ }\href {\doibase
  10.1103/PhysRevD.96.034025} {\bibfield  {journal} {\bibinfo  {journal} {Phys.
  Rev. D}\ }\textbf {\bibinfo {volume} {96}},\ \bibinfo {pages} {034025}
  (\bibinfo {year} {2017})},\ \Eprint {http://arxiv.org/abs/1705.01488}
  {arXiv:1705.01488 [hep-ph]} \BibitemShut {NoStop}%
\bibitem [{\citenamefont {Chambers}\ \emph {et~al.}(2017)\citenamefont
  {Chambers}, \citenamefont {Horsley}, \citenamefont {Nakamura}, \citenamefont
  {Perlt}, \citenamefont {Rakow}, \citenamefont {Schierholz}, \citenamefont
  {Schiller}, \citenamefont {Somfleth}, \citenamefont {Young},\ and\
  \citenamefont {Zanotti}}]{Chambers:2017dov}%
  \BibitemOpen
  \bibfield  {author} {\bibinfo {author} {\bibfnamefont {A.~J.}\ \bibnamefont
  {Chambers}}, \bibinfo {author} {\bibfnamefont {R.}~\bibnamefont {Horsley}},
  \bibinfo {author} {\bibfnamefont {Y.}~\bibnamefont {Nakamura}}, \bibinfo
  {author} {\bibfnamefont {H.}~\bibnamefont {Perlt}}, \bibinfo {author}
  {\bibfnamefont {P.~E.~L.}\ \bibnamefont {Rakow}}, \bibinfo {author}
  {\bibfnamefont {G.}~\bibnamefont {Schierholz}}, \bibinfo {author}
  {\bibfnamefont {A.}~\bibnamefont {Schiller}}, \bibinfo {author}
  {\bibfnamefont {K.}~\bibnamefont {Somfleth}}, \bibinfo {author}
  {\bibfnamefont {R.~D.}\ \bibnamefont {Young}}, \ and\ \bibinfo {author}
  {\bibfnamefont {J.~M.}\ \bibnamefont {Zanotti}},\ }\bibfield  {title}
  {\enquote {\bibinfo {title} {{Nucleon Structure Functions from Operator
  Product Expansion on the Lattice}},}\ }\href {\doibase
  10.1103/PhysRevLett.118.242001} {\bibfield  {journal} {\bibinfo  {journal}
  {Phys. Rev. Lett.}\ }\textbf {\bibinfo {volume} {118}},\ \bibinfo {pages}
  {242001} (\bibinfo {year} {2017})},\ \Eprint
  {http://arxiv.org/abs/1703.01153} {arXiv:1703.01153 [hep-lat]} \BibitemShut
  {NoStop}%
\bibitem [{\citenamefont {Davoudi}\ and\ \citenamefont
  {Savage}(2012)}]{Davoudi:2012ya}%
  \BibitemOpen
  \bibfield  {author} {\bibinfo {author} {\bibfnamefont {Zohreh}\ \bibnamefont
  {Davoudi}}\ and\ \bibinfo {author} {\bibfnamefont {Martin~J.}\ \bibnamefont
  {Savage}},\ }\bibfield  {title} {\enquote {\bibinfo {title} {{Restoration of
  Rotational Symmetry in the Continuum Limit of Lattice Field Theories}},}\
  }\href {\doibase 10.1103/PhysRevD.86.054505} {\bibfield  {journal} {\bibinfo
  {journal} {Phys. Rev. D}\ }\textbf {\bibinfo {volume} {86}},\ \bibinfo
  {pages} {054505} (\bibinfo {year} {2012})},\ \Eprint
  {http://arxiv.org/abs/1204.4146} {arXiv:1204.4146 [hep-lat]} \BibitemShut
  {NoStop}%
\bibitem [{\citenamefont {Asakawa}\ \emph {et~al.}(2001)\citenamefont
  {Asakawa}, \citenamefont {Hatsuda},\ and\ \citenamefont
  {Nakahara}}]{Asakawa:2000tr}%
  \BibitemOpen
  \bibfield  {author} {\bibinfo {author} {\bibfnamefont {M.}~\bibnamefont
  {Asakawa}}, \bibinfo {author} {\bibfnamefont {T.}~\bibnamefont {Hatsuda}}, \
  and\ \bibinfo {author} {\bibfnamefont {Y.}~\bibnamefont {Nakahara}},\
  }\bibfield  {title} {\enquote {\bibinfo {title} {{Maximum entropy analysis of
  the spectral functions in lattice QCD}},}\ }\href {\doibase
  10.1016/S0146-6410(01)00150-8} {\bibfield  {journal} {\bibinfo  {journal}
  {Prog. Part. Nucl. Phys.}\ }\textbf {\bibinfo {volume} {46}},\ \bibinfo
  {pages} {459--508} (\bibinfo {year} {2001})},\ \Eprint
  {http://arxiv.org/abs/hep-lat/0011040} {arXiv:hep-lat/0011040} \BibitemShut
  {NoStop}%
\bibitem [{\citenamefont {Aarts}\ and\ \citenamefont
  {Mart\'inez~Resco}(2002)}]{Aarts:2002cc}%
  \BibitemOpen
  \bibfield  {author} {\bibinfo {author} {\bibfnamefont {Gert}\ \bibnamefont
  {Aarts}}\ and\ \bibinfo {author} {\bibfnamefont {Jose~Maria}\ \bibnamefont
  {Mart\'inez~Resco}},\ }\bibfield  {title} {\enquote {\bibinfo {title}
  {{Transport coefficients, spectral functions and the lattice}},}\ }\href
  {\doibase 10.1088/1126-6708/2002/04/053} {\bibfield  {journal} {\bibinfo
  {journal} {JHEP}\ }\textbf {\bibinfo {volume} {04}},\ \bibinfo {pages} {053}
  (\bibinfo {year} {2002})},\ \Eprint {http://arxiv.org/abs/hep-ph/0203177}
  {arXiv:hep-ph/0203177} \BibitemShut {NoStop}%
\bibitem [{\citenamefont {Meyer}(2011)}]{Meyer:2011gj}%
  \BibitemOpen
  \bibfield  {author} {\bibinfo {author} {\bibfnamefont {Harvey~B.}\
  \bibnamefont {Meyer}},\ }\bibfield  {title} {\enquote {\bibinfo {title}
  {{Transport Properties of the Quark-Gluon Plasma: A Lattice QCD
  Perspective}},}\ }\href {\doibase 10.1140/epja/i2011-11086-3} {\bibfield
  {journal} {\bibinfo  {journal} {Eur. Phys. J. A}\ }\textbf {\bibinfo {volume}
  {47}},\ \bibinfo {pages} {86} (\bibinfo {year} {2011})},\ \Eprint
  {http://arxiv.org/abs/1104.3708} {arXiv:1104.3708 [hep-lat]} \BibitemShut
  {NoStop}%
\bibitem [{\citenamefont {Luscher}(2010)}]{Luscher:2009eq}%
  \BibitemOpen
  \bibfield  {author} {\bibinfo {author} {\bibfnamefont {Martin}\ \bibnamefont
  {Luscher}},\ }\bibfield  {title} {\enquote {\bibinfo {title} {{Trivializing
  maps, the Wilson flow and the HMC algorithm}},}\ }\href {\doibase
  10.1007/s00220-009-0953-7} {\bibfield  {journal} {\bibinfo  {journal}
  {Commun. Math. Phys.}\ }\textbf {\bibinfo {volume} {293}},\ \bibinfo {pages}
  {899--919} (\bibinfo {year} {2010})},\ \Eprint
  {http://arxiv.org/abs/0907.5491} {arXiv:0907.5491 [hep-lat]} \BibitemShut
  {NoStop}%
\bibitem [{\citenamefont {DeGrand}(2016)}]{degrandLatticeTestsStandard2016}%
  \BibitemOpen
  \bibfield  {author} {\bibinfo {author} {\bibfnamefont {Thomas}\ \bibnamefont
  {DeGrand}},\ }\bibfield  {title} {\enquote {\bibinfo {title} {Lattice tests
  of beyond standard model dynamics},}\ }\href {\doibase
  10.1103/RevModPhys.88.015001} {\bibfield  {journal} {\bibinfo  {journal}
  {Rev. Mod. Phys.}\ }\textbf {\bibinfo {volume} {88}},\ \bibinfo {pages}
  {015001} (\bibinfo {year} {2016})},\ \Eprint
  {http://arxiv.org/abs/1510.05018} {arXiv:1510.05018 [hep-ph]} \BibitemShut
  {NoStop}%
\bibitem [{\citenamefont {Drach}(2020)}]{Drach:2020qpj}%
  \BibitemOpen
  \bibfield  {author} {\bibinfo {author} {\bibfnamefont {V.}~\bibnamefont
  {Drach}},\ }\bibfield  {title} {\enquote {\bibinfo {title} {{Composite
  electroweak sectors on the lattice}},}\ }\href {\doibase 10.22323/1.363.0242}
  {\bibfield  {journal} {\bibinfo  {journal} {PoS}\ }\textbf {\bibinfo {volume}
  {LATTICE2019}},\ \bibinfo {pages} {242} (\bibinfo {year} {2020})},\ \Eprint
  {http://arxiv.org/abs/2005.01002} {arXiv:2005.01002 [hep-lat]} \BibitemShut
  {NoStop}%
\bibitem [{\citenamefont {Schaich}(2019)}]{Schaich:2018mmv}%
  \BibitemOpen
  \bibfield  {author} {\bibinfo {author} {\bibfnamefont {David}\ \bibnamefont
  {Schaich}},\ }\bibfield  {title} {\enquote {\bibinfo {title} {{Progress and
  prospects of lattice supersymmetry}},}\ }\href {\doibase 10.22323/1.334.0005}
  {\bibfield  {journal} {\bibinfo  {journal} {PoS}\ }\textbf {\bibinfo {volume}
  {LATTICE2018}},\ \bibinfo {pages} {005} (\bibinfo {year} {2019})},\ \Eprint
  {http://arxiv.org/abs/1810.09282} {arXiv:1810.09282 [hep-lat]} \BibitemShut
  {NoStop}%
\bibitem [{\citenamefont {Hern\'andez}\ and\ \citenamefont
  {Romero-L\'opez}(2021)}]{Hernandez:2020tbc}%
  \BibitemOpen
  \bibfield  {author} {\bibinfo {author} {\bibfnamefont {Pilar}\ \bibnamefont
  {Hern\'andez}}\ and\ \bibinfo {author} {\bibfnamefont {Fernando}\
  \bibnamefont {Romero-L\'opez}},\ }\bibfield  {title} {\enquote {\bibinfo
  {title} {{The large $N_{c}$ limit of QCD on the lattice}},}\ }\href {\doibase
  10.1140/epja/s10050-021-00374-2} {\bibfield  {journal} {\bibinfo  {journal}
  {Eur. Phys. J. A}\ }\textbf {\bibinfo {volume} {57}},\ \bibinfo {pages} {52}
  (\bibinfo {year} {2021})},\ \Eprint {http://arxiv.org/abs/2012.03331}
  {arXiv:2012.03331 [hep-lat]} \BibitemShut {NoStop}%
\bibitem [{\citenamefont {Lucini}\ and\ \citenamefont
  {Panero}(2013)}]{Lucini:2012gg}%
  \BibitemOpen
  \bibfield  {author} {\bibinfo {author} {\bibfnamefont {Biagio}\ \bibnamefont
  {Lucini}}\ and\ \bibinfo {author} {\bibfnamefont {Marco}\ \bibnamefont
  {Panero}},\ }\bibfield  {title} {\enquote {\bibinfo {title} {{SU(N) gauge
  theories at large N}},}\ }\href {\doibase 10.1016/j.physrep.2013.01.001}
  {\bibfield  {journal} {\bibinfo  {journal} {Phys. Rept.}\ }\textbf {\bibinfo
  {volume} {526}},\ \bibinfo {pages} {93--163} (\bibinfo {year} {2013})},\
  \Eprint {http://arxiv.org/abs/1210.4997} {arXiv:1210.4997 [hep-th]}
  \BibitemShut {NoStop}%
\bibitem [{\citenamefont {Greensite}(2003)}]{Greensite:2003bk}%
  \BibitemOpen
  \bibfield  {author} {\bibinfo {author} {\bibfnamefont {J.}~\bibnamefont
  {Greensite}},\ }\bibfield  {title} {\enquote {\bibinfo {title} {{The
  Confinement problem in lattice gauge theory}},}\ }\href {\doibase
  10.1016/S0146-6410(03)90012-3} {\bibfield  {journal} {\bibinfo  {journal}
  {Prog. Part. Nucl. Phys.}\ }\textbf {\bibinfo {volume} {51}},\ \bibinfo
  {pages} {1} (\bibinfo {year} {2003})},\ \Eprint
  {http://arxiv.org/abs/hep-lat/0301023} {arXiv:hep-lat/0301023} \BibitemShut
  {NoStop}%
\bibitem [{\citenamefont {Ripka}(2004)}]{Ripka:2003vv}%
  \BibitemOpen
  \bibfield  {author} {\bibinfo {author} {\bibfnamefont {Georges}\ \bibnamefont
  {Ripka}},\ }\href {\doibase 10.1007/b94800} {\emph {\bibinfo {title} {{Dual
  superconductor models of color confinement}}}},\ Vol.\ \bibinfo {volume}
  {639}\ (\bibinfo {year} {2004})\ \Eprint
  {http://arxiv.org/abs/hep-ph/0310102} {arXiv:hep-ph/0310102} \BibitemShut
  {NoStop}%
\bibitem [{\citenamefont {Mathur}\ and\ \citenamefont
  {Sreeraj}(2016)}]{Mathur:2016cko}%
  \BibitemOpen
  \bibfield  {author} {\bibinfo {author} {\bibfnamefont {Manu}\ \bibnamefont
  {Mathur}}\ and\ \bibinfo {author} {\bibfnamefont {T.~P.}\ \bibnamefont
  {Sreeraj}},\ }\bibfield  {title} {\enquote {\bibinfo {title} {{Lattice Gauge
  Theories and Spin Models}},}\ }\href {\doibase 10.1103/PhysRevD.94.085029}
  {\bibfield  {journal} {\bibinfo  {journal} {Phys. Rev. D}\ }\textbf {\bibinfo
  {volume} {94}},\ \bibinfo {pages} {085029} (\bibinfo {year} {2016})},\
  \Eprint {http://arxiv.org/abs/1604.00315} {arXiv:1604.00315 [hep-lat]}
  \BibitemShut {NoStop}%
\bibitem [{\citenamefont {Wang}(2017)}]{Wang2017}%
  \BibitemOpen
  \bibfield  {author} {\bibinfo {author} {\bibfnamefont {Lei}\ \bibnamefont
  {Wang}},\ }\bibfield  {title} {\enquote {\bibinfo {title} {{Exploring cluster
  Monte Carlo updates with Boltzmann machines}},}\ }\href {\doibase
  10.1103/PhysRevE.96.051301} {\bibfield  {journal} {\bibinfo  {journal} {Phys.
  Rev. E}\ }\textbf {\bibinfo {volume} {96}},\ \bibinfo {pages} {051301}
  (\bibinfo {year} {2017})}\BibitemShut {NoStop}%
\bibitem [{\citenamefont {Huang}\ and\ \citenamefont
  {Wang}(2017)}]{Huang:2017}%
  \BibitemOpen
  \bibfield  {author} {\bibinfo {author} {\bibfnamefont {Li}~\bibnamefont
  {Huang}}\ and\ \bibinfo {author} {\bibfnamefont {Lei}\ \bibnamefont {Wang}},\
  }\bibfield  {title} {\enquote {\bibinfo {title} {{Accelerated Monte Carlo
  simulations with restricted Boltzmann machines}},}\ }\href {\doibase
  10.1103/physrevb.95.035105} {\bibfield  {journal} {\bibinfo  {journal}
  {Physical Review B}\ }\textbf {\bibinfo {volume} {95}},\ \bibinfo {pages}
  {--} (\bibinfo {year} {2017})}\BibitemShut {NoStop}%
\bibitem [{\citenamefont {Song}\ \emph {et~al.}(2017)\citenamefont {Song},
  \citenamefont {Zhao},\ and\ \citenamefont {Ermon}}]{song2017nice}%
  \BibitemOpen
  \bibfield  {author} {\bibinfo {author} {\bibfnamefont {Jiaming}\ \bibnamefont
  {Song}}, \bibinfo {author} {\bibfnamefont {Shengjia}\ \bibnamefont {Zhao}}, \
  and\ \bibinfo {author} {\bibfnamefont {Stefano}\ \bibnamefont {Ermon}},\
  }\bibfield  {title} {\enquote {\bibinfo {title} {{A-NICE-MC: Adversarial
  training for MCMC}},}\ }in\ \href@noop {} {\emph {\bibinfo {booktitle}
  {Advances in Neural Information Processing Systems}}}\ (\bibinfo {year}
  {2017})\ pp.\ \bibinfo {pages} {5140--5150}\BibitemShut {NoStop}%
\bibitem [{\citenamefont {Li}\ and\ \citenamefont
  {Wang}(2018)}]{LiWang2018NNRG}%
  \BibitemOpen
  \bibfield  {author} {\bibinfo {author} {\bibfnamefont {Shuo-Hui}\
  \bibnamefont {Li}}\ and\ \bibinfo {author} {\bibfnamefont {Lei}\ \bibnamefont
  {Wang}},\ }\bibfield  {title} {\enquote {\bibinfo {title} {{Neural Network
  Renormalization Group}},}\ }\href {\doibase 10.1103/PhysRevLett.121.260601}
  {\bibfield  {journal} {\bibinfo  {journal} {Phys. Rev. Lett.}\ }\textbf
  {\bibinfo {volume} {121}},\ \bibinfo {pages} {260601} (\bibinfo {year}
  {2018})}\BibitemShut {NoStop}%
\bibitem [{\citenamefont {Levy}\ \emph {et~al.}(2018)\citenamefont {Levy},
  \citenamefont {Hoffman},\ and\ \citenamefont
  {Sohl-Dickstein}}]{levy2018generalizing}%
  \BibitemOpen
  \bibfield  {author} {\bibinfo {author} {\bibfnamefont {Daniel}\ \bibnamefont
  {Levy}}, \bibinfo {author} {\bibfnamefont {Matthew~D.}\ \bibnamefont
  {Hoffman}}, \ and\ \bibinfo {author} {\bibfnamefont {Jascha}\ \bibnamefont
  {Sohl-Dickstein}},\ }\bibfield  {title} {\enquote {\bibinfo {title}
  {Generalizing hamiltonian monte carlo with neural networks},}\ }\href@noop {}
  {\  (\bibinfo {year} {2018})},\ \Eprint {http://arxiv.org/abs/1711.09268}
  {arXiv:1711.09268 [stat.ML]} \BibitemShut {NoStop}%
\bibitem [{\citenamefont {Albergo}\ \emph {et~al.}(2019)\citenamefont
  {Albergo}, \citenamefont {Kanwar},\ and\ \citenamefont
  {Shanahan}}]{Albergo:2019eim}%
  \BibitemOpen
  \bibfield  {author} {\bibinfo {author} {\bibfnamefont {M.~S.}\ \bibnamefont
  {Albergo}}, \bibinfo {author} {\bibfnamefont {G.}~\bibnamefont {Kanwar}}, \
  and\ \bibinfo {author} {\bibfnamefont {P.~E.}\ \bibnamefont {Shanahan}},\
  }\bibfield  {title} {\enquote {\bibinfo {title} {{Flow-based generative
  models for Markov chain Monte Carlo in lattice field theory}},}\ }\href
  {\doibase 10.1103/PhysRevD.100.034515} {\bibfield  {journal} {\bibinfo
  {journal} {Phys. Rev. D}\ }\textbf {\bibinfo {volume} {100}},\ \bibinfo
  {pages} {034515} (\bibinfo {year} {2019})},\ \Eprint
  {http://arxiv.org/abs/1904.12072} {arXiv:1904.12072 [hep-lat]} \BibitemShut
  {NoStop}%
\bibitem [{\citenamefont {Wu}\ \emph {et~al.}(2019)\citenamefont {Wu},
  \citenamefont {Wang},\ and\ \citenamefont {Zhang}}]{Wu:2019}%
  \BibitemOpen
  \bibfield  {author} {\bibinfo {author} {\bibfnamefont {Dian}\ \bibnamefont
  {Wu}}, \bibinfo {author} {\bibfnamefont {Lei}\ \bibnamefont {Wang}}, \ and\
  \bibinfo {author} {\bibfnamefont {Pan}\ \bibnamefont {Zhang}},\ }\bibfield
  {title} {\enquote {\bibinfo {title} {{Solving Statistical Mechanics Using
  Variational Autoregressive Networks}},}\ }\href {\doibase
  10.1103/PhysRevLett.122.080602} {\bibfield  {journal} {\bibinfo  {journal}
  {Phys. Rev. Lett.}\ }\textbf {\bibinfo {volume} {122}},\ \bibinfo {pages}
  {080602} (\bibinfo {year} {2019})}\BibitemShut {NoStop}%
\bibitem [{\citenamefont {Nicoli}\ \emph {et~al.}(2020)\citenamefont {Nicoli},
  \citenamefont {Nakajima}, \citenamefont {Strodthoff}, \citenamefont {Samek},
  \citenamefont {Müller},\ and\ \citenamefont {Kessel}}]{Nicoli:2020evf}%
  \BibitemOpen
  \bibfield  {author} {\bibinfo {author} {\bibfnamefont {Kim~A.}\ \bibnamefont
  {Nicoli}}, \bibinfo {author} {\bibfnamefont {Shinichi}\ \bibnamefont
  {Nakajima}}, \bibinfo {author} {\bibfnamefont {Nils}\ \bibnamefont
  {Strodthoff}}, \bibinfo {author} {\bibfnamefont {Wojciech}\ \bibnamefont
  {Samek}}, \bibinfo {author} {\bibfnamefont {Klaus-Robert}\ \bibnamefont
  {Müller}}, \ and\ \bibinfo {author} {\bibfnamefont {Pan}\ \bibnamefont
  {Kessel}},\ }\bibfield  {title} {\enquote {\bibinfo {title} {{Asymptotically
  unbiased estimation of physical observables with neural samplers}},}\ }\href
  {\doibase 10.1103/PhysRevE.101.023304} {\bibfield  {journal} {\bibinfo
  {journal} {Phys. Rev. E}\ }\textbf {\bibinfo {volume} {101}},\ \bibinfo
  {pages} {023304} (\bibinfo {year} {2020})},\ \Eprint
  {http://arxiv.org/abs/1910.13496} {arXiv:1910.13496 [cond-mat.stat-mech]}
  \BibitemShut {NoStop}%
\bibitem [{\citenamefont {Nicoli}\ \emph {et~al.}(2021)\citenamefont {Nicoli},
  \citenamefont {Anders}, \citenamefont {Funcke}, \citenamefont {Hartung},
  \citenamefont {Jansen}, \citenamefont {Kessel}, \citenamefont {Nakajima},\
  and\ \citenamefont {Stornati}}]{Nicoli:2020njz}%
  \BibitemOpen
  \bibfield  {author} {\bibinfo {author} {\bibfnamefont {Kim~A.}\ \bibnamefont
  {Nicoli}}, \bibinfo {author} {\bibfnamefont {Christopher~J.}\ \bibnamefont
  {Anders}}, \bibinfo {author} {\bibfnamefont {Lena}\ \bibnamefont {Funcke}},
  \bibinfo {author} {\bibfnamefont {Tobias}\ \bibnamefont {Hartung}}, \bibinfo
  {author} {\bibfnamefont {Karl}\ \bibnamefont {Jansen}}, \bibinfo {author}
  {\bibfnamefont {Pan}\ \bibnamefont {Kessel}}, \bibinfo {author}
  {\bibfnamefont {Shinichi}\ \bibnamefont {Nakajima}}, \ and\ \bibinfo {author}
  {\bibfnamefont {Paolo}\ \bibnamefont {Stornati}},\ }\bibfield  {title}
  {\enquote {\bibinfo {title} {{Estimation of Thermodynamic Observables in
  Lattice Field Theories with Deep Generative Models}},}\ }\href {\doibase
  10.1103/PhysRevLett.126.032001} {\bibfield  {journal} {\bibinfo  {journal}
  {Phys. Rev. Lett.}\ }\textbf {\bibinfo {volume} {126}},\ \bibinfo {pages}
  {032001} (\bibinfo {year} {2021})},\ \Eprint
  {http://arxiv.org/abs/2007.07115} {arXiv:2007.07115 [hep-lat]} \BibitemShut
  {NoStop}%
\bibitem [{\citenamefont {Pawlowski}\ and\ \citenamefont
  {Urban}(2020)}]{Pawlowski:2018qxs}%
  \BibitemOpen
  \bibfield  {author} {\bibinfo {author} {\bibfnamefont {Jan~M.}\ \bibnamefont
  {Pawlowski}}\ and\ \bibinfo {author} {\bibfnamefont {Julian~M.}\ \bibnamefont
  {Urban}},\ }\bibfield  {title} {\enquote {\bibinfo {title} {{Reducing
  Autocorrelation Times in Lattice Simulations with Generative Adversarial
  Networks}},}\ }\href {\doibase 10.1088/2632-2153/abae73} {\bibfield
  {journal} {\bibinfo  {journal} {Mach. Learn. Sci. Tech.}\ }\textbf {\bibinfo
  {volume} {1}},\ \bibinfo {pages} {045011} (\bibinfo {year} {2020})},\ \Eprint
  {http://arxiv.org/abs/1811.03533} {arXiv:1811.03533 [hep-lat]} \BibitemShut
  {NoStop}%
\bibitem [{\citenamefont {Kanwar}\ \emph {et~al.}(2020)\citenamefont {Kanwar},
  \citenamefont {Albergo}, \citenamefont {Boyda}, \citenamefont {Cranmer},
  \citenamefont {Hackett}, \citenamefont {Racani\`ere}, \citenamefont
  {Rezende},\ and\ \citenamefont {Shanahan}}]{Kanwar:2020xzo}%
  \BibitemOpen
  \bibfield  {author} {\bibinfo {author} {\bibfnamefont {Gurtej}\ \bibnamefont
  {Kanwar}}, \bibinfo {author} {\bibfnamefont {Michael~S.}\ \bibnamefont
  {Albergo}}, \bibinfo {author} {\bibfnamefont {Denis}\ \bibnamefont {Boyda}},
  \bibinfo {author} {\bibfnamefont {Kyle}\ \bibnamefont {Cranmer}}, \bibinfo
  {author} {\bibfnamefont {Daniel~C.}\ \bibnamefont {Hackett}}, \bibinfo
  {author} {\bibfnamefont {S\'ebastien}\ \bibnamefont {Racani\`ere}}, \bibinfo
  {author} {\bibfnamefont {Danilo~Jimenez}\ \bibnamefont {Rezende}}, \ and\
  \bibinfo {author} {\bibfnamefont {Phiala~E.}\ \bibnamefont {Shanahan}},\
  }\bibfield  {title} {\enquote {\bibinfo {title} {{Equivariant flow-based
  sampling for lattice gauge theory}},}\ }\href {\doibase
  10.1103/PhysRevLett.125.121601} {\bibfield  {journal} {\bibinfo  {journal}
  {Phys. Rev. Lett.}\ }\textbf {\bibinfo {volume} {125}},\ \bibinfo {pages}
  {121601} (\bibinfo {year} {2020})},\ \Eprint
  {http://arxiv.org/abs/2003.06413} {arXiv:2003.06413 [hep-lat]} \BibitemShut
  {NoStop}%
\bibitem [{\citenamefont {Boyda}\ \emph
  {et~al.}(2021{\natexlab{a}})\citenamefont {Boyda}, \citenamefont {Kanwar},
  \citenamefont {Racani\`ere}, \citenamefont {Rezende}, \citenamefont
  {Albergo}, \citenamefont {Cranmer}, \citenamefont {Hackett},\ and\
  \citenamefont {Shanahan}}]{Boyda:2020hsi}%
  \BibitemOpen
  \bibfield  {author} {\bibinfo {author} {\bibfnamefont {Denis}\ \bibnamefont
  {Boyda}}, \bibinfo {author} {\bibfnamefont {Gurtej}\ \bibnamefont {Kanwar}},
  \bibinfo {author} {\bibfnamefont {S\'ebastien}\ \bibnamefont {Racani\`ere}},
  \bibinfo {author} {\bibfnamefont {Danilo~Jimenez}\ \bibnamefont {Rezende}},
  \bibinfo {author} {\bibfnamefont {Michael~S.}\ \bibnamefont {Albergo}},
  \bibinfo {author} {\bibfnamefont {Kyle}\ \bibnamefont {Cranmer}}, \bibinfo
  {author} {\bibfnamefont {Daniel~C.}\ \bibnamefont {Hackett}}, \ and\ \bibinfo
  {author} {\bibfnamefont {Phiala~E.}\ \bibnamefont {Shanahan}},\ }\bibfield
  {title} {\enquote {\bibinfo {title} {{Sampling using $SU(N)$ gauge
  equivariant flows}},}\ }\href {\doibase 10.1103/PhysRevD.103.074504}
  {\bibfield  {journal} {\bibinfo  {journal} {Phys. Rev. D}\ }\textbf {\bibinfo
  {volume} {103}},\ \bibinfo {pages} {074504} (\bibinfo {year}
  {2021}{\natexlab{a}})},\ \Eprint {http://arxiv.org/abs/2008.05456}
  {arXiv:2008.05456 [hep-lat]} \BibitemShut {NoStop}%
\bibitem [{\citenamefont {Albergo}\ \emph {et~al.}(2021)\citenamefont
  {Albergo}, \citenamefont {Kanwar}, \citenamefont {Racani\`ere}, \citenamefont
  {Rezende}, \citenamefont {Urban}, \citenamefont {Boyda}, \citenamefont
  {Cranmer}, \citenamefont {Hackett},\ and\ \citenamefont
  {Shanahan}}]{Albergo:2021bna}%
  \BibitemOpen
  \bibfield  {author} {\bibinfo {author} {\bibfnamefont {Michael~S.}\
  \bibnamefont {Albergo}}, \bibinfo {author} {\bibfnamefont {Gurtej}\
  \bibnamefont {Kanwar}}, \bibinfo {author} {\bibfnamefont {S\'ebastien}\
  \bibnamefont {Racani\`ere}}, \bibinfo {author} {\bibfnamefont {Danilo~J.}\
  \bibnamefont {Rezende}}, \bibinfo {author} {\bibfnamefont {Julian~M.}\
  \bibnamefont {Urban}}, \bibinfo {author} {\bibfnamefont {Denis}\ \bibnamefont
  {Boyda}}, \bibinfo {author} {\bibfnamefont {Kyle}\ \bibnamefont {Cranmer}},
  \bibinfo {author} {\bibfnamefont {Daniel~C.}\ \bibnamefont {Hackett}}, \ and\
  \bibinfo {author} {\bibfnamefont {Phiala~E.}\ \bibnamefont {Shanahan}},\
  }\href@noop {} {\enquote {\bibinfo {title} {{Flow-based sampling for
  fermionic lattice field theories}},}\ } (\bibinfo {year} {2021}),\ \Eprint
  {http://arxiv.org/abs/2106.05934} {arXiv:2106.05934 [hep-lat]} \BibitemShut
  {NoStop}%
\bibitem [{\citenamefont {Hackett}\ \emph {et~al.}(2021)\citenamefont
  {Hackett}, \citenamefont {Hsieh}, \citenamefont {Albergo}, \citenamefont
  {Boyda}, \citenamefont {Chen}, \citenamefont {Chen}, \citenamefont {Cranmer},
  \citenamefont {Kanwar},\ and\ \citenamefont {Shanahan}}]{Hackett:2021idh}%
  \BibitemOpen
  \bibfield  {author} {\bibinfo {author} {\bibfnamefont {Daniel~C.}\
  \bibnamefont {Hackett}}, \bibinfo {author} {\bibfnamefont {Chung-Chun}\
  \bibnamefont {Hsieh}}, \bibinfo {author} {\bibfnamefont {Michael~S.}\
  \bibnamefont {Albergo}}, \bibinfo {author} {\bibfnamefont {Denis}\
  \bibnamefont {Boyda}}, \bibinfo {author} {\bibfnamefont {Jiunn-Wei}\
  \bibnamefont {Chen}}, \bibinfo {author} {\bibfnamefont {Kai-Feng}\
  \bibnamefont {Chen}}, \bibinfo {author} {\bibfnamefont {Kyle}\ \bibnamefont
  {Cranmer}}, \bibinfo {author} {\bibfnamefont {Gurtej}\ \bibnamefont
  {Kanwar}}, \ and\ \bibinfo {author} {\bibfnamefont {Phiala~E.}\ \bibnamefont
  {Shanahan}},\ }\bibfield  {title} {\enquote {\bibinfo {title} {{Flow-based
  sampling for multimodal distributions in lattice field theory}},}\
  }\href@noop {} {\  (\bibinfo {year} {2021})},\ \Eprint
  {http://arxiv.org/abs/2107.00734} {arXiv:2107.00734 [hep-lat]} \BibitemShut
  {NoStop}%
\bibitem [{\citenamefont {Gabri\'e}\ \emph {et~al.}(2021)\citenamefont
  {Gabri\'e}, \citenamefont {Rotskoff},\ and\ \citenamefont
  {Vanden-Eijnden}}]{Gabrie:2021tlu}%
  \BibitemOpen
  \bibfield  {author} {\bibinfo {author} {\bibfnamefont {Marylou}\ \bibnamefont
  {Gabri\'e}}, \bibinfo {author} {\bibfnamefont {Grant~M.}\ \bibnamefont
  {Rotskoff}}, \ and\ \bibinfo {author} {\bibfnamefont {Eric}\ \bibnamefont
  {Vanden-Eijnden}},\ }\bibfield  {title} {\enquote {\bibinfo {title}
  {{Adaptive Monte Carlo augmented with normalizing flows}},}\ }\href@noop {}
  {\  (\bibinfo {year} {2021})},\ \Eprint {http://arxiv.org/abs/2105.12603}
  {arXiv:2105.12603 [physics.data-an]} \BibitemShut {NoStop}%
\bibitem [{\citenamefont {Del~Debbio}\ \emph {et~al.}(2021)\citenamefont
  {Del~Debbio}, \citenamefont {Rossney},\ and\ \citenamefont
  {Wilson}}]{DelDebbio:2021qwf}%
  \BibitemOpen
  \bibfield  {author} {\bibinfo {author} {\bibfnamefont {Luigi}\ \bibnamefont
  {Del~Debbio}}, \bibinfo {author} {\bibfnamefont {Joe~Marsh}\ \bibnamefont
  {Rossney}}, \ and\ \bibinfo {author} {\bibfnamefont {Michael}\ \bibnamefont
  {Wilson}},\ }\bibfield  {title} {\enquote {\bibinfo {title} {{Efficient
  Modelling of Trivializing Maps for Lattice $\phi^4$ Theory Using Normalizing
  Flows: A First Look at Scalability}},}\ }\href@noop {} {\  (\bibinfo {year}
  {2021})},\ \Eprint {http://arxiv.org/abs/2105.12481} {arXiv:2105.12481
  [hep-lat]} \BibitemShut {NoStop}%
\bibitem [{\citenamefont {Wu}\ \emph {et~al.}(2021)\citenamefont {Wu},
  \citenamefont {Rossi},\ and\ \citenamefont {Carleo}}]{Wu:2021tfb}%
  \BibitemOpen
  \bibfield  {author} {\bibinfo {author} {\bibfnamefont {Dian}\ \bibnamefont
  {Wu}}, \bibinfo {author} {\bibfnamefont {Riccardo}\ \bibnamefont {Rossi}}, \
  and\ \bibinfo {author} {\bibfnamefont {Giuseppe}\ \bibnamefont {Carleo}},\
  }\bibfield  {title} {\enquote {\bibinfo {title} {{Unbiased Monte Carlo
  Cluster Updates with Autoregressive Neural Networks}},}\ }\href@noop {} {\
  (\bibinfo {year} {2021})},\ \Eprint {http://arxiv.org/abs/2105.05650}
  {arXiv:2105.05650 [cond-mat.stat-mech]} \BibitemShut {NoStop}%
\bibitem [{\citenamefont {Foreman}\ \emph
  {et~al.}(2021{\natexlab{a}})\citenamefont {Foreman}, \citenamefont {Jin},\
  and\ \citenamefont {Osborn}}]{Foreman:2021ixr}%
  \BibitemOpen
  \bibfield  {author} {\bibinfo {author} {\bibfnamefont {Sam}\ \bibnamefont
  {Foreman}}, \bibinfo {author} {\bibfnamefont {Xiao-Yong}\ \bibnamefont
  {Jin}}, \ and\ \bibinfo {author} {\bibfnamefont {James~C.}\ \bibnamefont
  {Osborn}},\ }\bibfield  {title} {\enquote {\bibinfo {title} {{Deep Learning
  Hamiltonian Monte Carlo}},}\ }in\ \href@noop {} {\emph {\bibinfo {booktitle}
  {{9th International Conference on Learning Representations}}}}\ (\bibinfo
  {year} {2021})\ \Eprint {http://arxiv.org/abs/2105.03418} {arXiv:2105.03418
  [hep-lat]} \BibitemShut {NoStop}%
\bibitem [{\citenamefont {Foreman}\ \emph
  {et~al.}(2021{\natexlab{b}})\citenamefont {Foreman}, \citenamefont
  {Izubuchi}, \citenamefont {Jin}, \citenamefont {Jin}, \citenamefont
  {Osborn},\ and\ \citenamefont {Tomiya}}]{Foreman:2021ljl}%
  \BibitemOpen
  \bibfield  {author} {\bibinfo {author} {\bibfnamefont {Sam}\ \bibnamefont
  {Foreman}}, \bibinfo {author} {\bibfnamefont {Taku}\ \bibnamefont
  {Izubuchi}}, \bibinfo {author} {\bibfnamefont {Luchang}\ \bibnamefont {Jin}},
  \bibinfo {author} {\bibfnamefont {Xiao-Yong}\ \bibnamefont {Jin}}, \bibinfo
  {author} {\bibfnamefont {James~C.}\ \bibnamefont {Osborn}}, \ and\ \bibinfo
  {author} {\bibfnamefont {Akio}\ \bibnamefont {Tomiya}},\ }\bibfield  {title}
  {\enquote {\bibinfo {title} {{HMC with Normalizing Flows}},}\ }in\ \href@noop
  {} {\emph {\bibinfo {booktitle} {{38th International Symposium on Lattice
  Field Theory}}}}\ (\bibinfo {year} {2021})\ \Eprint
  {http://arxiv.org/abs/2112.01586} {arXiv:2112.01586 [cs.LG]} \BibitemShut
  {NoStop}%
\bibitem [{\citenamefont {Foreman}\ \emph
  {et~al.}(2021{\natexlab{c}})\citenamefont {Foreman}, \citenamefont {Jin},\
  and\ \citenamefont {Osborn}}]{Foreman:2021rhs}%
  \BibitemOpen
  \bibfield  {author} {\bibinfo {author} {\bibfnamefont {Sam}\ \bibnamefont
  {Foreman}}, \bibinfo {author} {\bibfnamefont {Xiao-Yong}\ \bibnamefont
  {Jin}}, \ and\ \bibinfo {author} {\bibfnamefont {James~C.}\ \bibnamefont
  {Osborn}},\ }\bibfield  {title} {\enquote {\bibinfo {title} {{LeapfrogLayers:
  A Trainable Framework for Effective Topological Sampling}},}\ }in\ \href@noop
  {} {\emph {\bibinfo {booktitle} {{38th International Symposium on Lattice
  Field Theory}}}}\ (\bibinfo {year} {2021})\ \Eprint
  {http://arxiv.org/abs/2112.01582} {arXiv:2112.01582 [hep-lat]} \BibitemShut
  {NoStop}%
\bibitem [{\citenamefont {Jin}(2022)}]{Jin:2022bgq}%
  \BibitemOpen
  \bibfield  {author} {\bibinfo {author} {\bibfnamefont {Xiao-Yong}\
  \bibnamefont {Jin}},\ }\bibfield  {title} {\enquote {\bibinfo {title}
  {{Neural Network Field Transformation and Its Application in HMC}},}\ }in\
  \href@noop {} {\emph {\bibinfo {booktitle} {{38th International Symposium on
  Lattice Field Theory}}}}\ (\bibinfo {year} {2022})\ \Eprint
  {http://arxiv.org/abs/2201.01862} {arXiv:2201.01862 [hep-lat]} \BibitemShut
  {NoStop}%
\bibitem [{\citenamefont {Finkenrath}(2022)}]{Finkenrath:2022ogg}%
  \BibitemOpen
  \bibfield  {author} {\bibinfo {author} {\bibfnamefont {Jacob}\ \bibnamefont
  {Finkenrath}},\ }\bibfield  {title} {\enquote {\bibinfo {title} {{Tackling
  critical slowing down using global correction steps with equivariant flows:
  the case of the Schwinger model}},}\ }\href@noop {} {\  (\bibinfo {year}
  {2022})},\ \Eprint {http://arxiv.org/abs/2201.02216} {arXiv:2201.02216
  [hep-lat]} \BibitemShut {NoStop}%
\bibitem [{\citenamefont {Bachtis}\ \emph {et~al.}(2020)\citenamefont
  {Bachtis}, \citenamefont {Aarts},\ and\ \citenamefont
  {Lucini}}]{Bachtis:2020dmf}%
  \BibitemOpen
  \bibfield  {author} {\bibinfo {author} {\bibfnamefont {Dimitrios}\
  \bibnamefont {Bachtis}}, \bibinfo {author} {\bibfnamefont {Gert}\
  \bibnamefont {Aarts}}, \ and\ \bibinfo {author} {\bibfnamefont {Biagio}\
  \bibnamefont {Lucini}},\ }\bibfield  {title} {\enquote {\bibinfo {title}
  {{Extending machine learning classification capabilities with histogram
  reweighting}},}\ }\href {\doibase 10.1103/PhysRevE.102.033303} {\bibfield
  {journal} {\bibinfo  {journal} {Phys. Rev. E}\ }\textbf {\bibinfo {volume}
  {102}},\ \bibinfo {pages} {033303} (\bibinfo {year} {2020})},\ \Eprint
  {http://arxiv.org/abs/2004.14341} {arXiv:2004.14341 [cond-mat.stat-mech]}
  \BibitemShut {NoStop}%
\bibitem [{\citenamefont {de~Haan}\ \emph {et~al.}(2021)\citenamefont
  {de~Haan}, \citenamefont {Rainone}, \citenamefont {Cheng},\ and\
  \citenamefont {Bondesan}}]{deHaan:2021erb}%
  \BibitemOpen
  \bibfield  {author} {\bibinfo {author} {\bibfnamefont {Pim}\ \bibnamefont
  {de~Haan}}, \bibinfo {author} {\bibfnamefont {Corrado}\ \bibnamefont
  {Rainone}}, \bibinfo {author} {\bibfnamefont {Miranda C.~N.}\ \bibnamefont
  {Cheng}}, \ and\ \bibinfo {author} {\bibfnamefont {Roberto}\ \bibnamefont
  {Bondesan}},\ }\bibfield  {title} {\enquote {\bibinfo {title} {{Scaling Up
  Machine Learning For Quantum Field Theory with Equivariant Continuous
  Flows}},}\ }\href@noop {} {\  (\bibinfo {year} {2021})},\ \Eprint
  {http://arxiv.org/abs/2110.02673} {arXiv:2110.02673 [cs.LG]} \BibitemShut
  {NoStop}%
\bibitem [{\citenamefont {Cossu}\ \emph {et~al.}(2019)\citenamefont {Cossu},
  \citenamefont {Del~Debbio}, \citenamefont {Giani}, \citenamefont {Khamseh},\
  and\ \citenamefont {Wilson}}]{Cossu:2018pxj}%
  \BibitemOpen
  \bibfield  {author} {\bibinfo {author} {\bibfnamefont {Guido}\ \bibnamefont
  {Cossu}}, \bibinfo {author} {\bibfnamefont {Luigi}\ \bibnamefont
  {Del~Debbio}}, \bibinfo {author} {\bibfnamefont {Tommaso}\ \bibnamefont
  {Giani}}, \bibinfo {author} {\bibfnamefont {Ava}\ \bibnamefont {Khamseh}}, \
  and\ \bibinfo {author} {\bibfnamefont {Michael}\ \bibnamefont {Wilson}},\
  }\bibfield  {title} {\enquote {\bibinfo {title} {{Machine learning
  determination of dynamical parameters: The Ising model case}},}\ }\href
  {\doibase 10.1103/PhysRevB.100.064304} {\bibfield  {journal} {\bibinfo
  {journal} {Phys. Rev. B}\ }\textbf {\bibinfo {volume} {100}},\ \bibinfo
  {pages} {064304} (\bibinfo {year} {2019})},\ \Eprint
  {http://arxiv.org/abs/1810.11503} {arXiv:1810.11503 [physics.comp-ph]}
  \BibitemShut {NoStop}%
\bibitem [{\citenamefont {Tanaka}\ and\ \citenamefont
  {Tomiya}(2017)}]{Tanaka:2017niz}%
  \BibitemOpen
  \bibfield  {author} {\bibinfo {author} {\bibfnamefont {Akinori}\ \bibnamefont
  {Tanaka}}\ and\ \bibinfo {author} {\bibfnamefont {Akio}\ \bibnamefont
  {Tomiya}},\ }\bibfield  {title} {\enquote {\bibinfo {title} {{Towards
  reduction of autocorrelation in HMC by machine learning}},}\ }\href@noop {}
  {\  (\bibinfo {year} {2017})},\ \Eprint {http://arxiv.org/abs/1712.03893}
  {arXiv:1712.03893 [hep-lat]} \BibitemShut {NoStop}%
\bibitem [{\citenamefont {Tomiya}\ and\ \citenamefont
  {Nagai}(2021)}]{Tomiya:2021ywc}%
  \BibitemOpen
  \bibfield  {author} {\bibinfo {author} {\bibfnamefont {Akio}\ \bibnamefont
  {Tomiya}}\ and\ \bibinfo {author} {\bibfnamefont {Yuki}\ \bibnamefont
  {Nagai}},\ }\bibfield  {title} {\enquote {\bibinfo {title} {{Gauge covariant
  neural network for 4 dimensional non-abelian gauge theory}},}\ }\href@noop {}
  {\  (\bibinfo {year} {2021})},\ \Eprint {http://arxiv.org/abs/2103.11965}
  {arXiv:2103.11965 [hep-lat]} \BibitemShut {NoStop}%
\bibitem [{\citenamefont {Bachtis}\ \emph
  {et~al.}(2021{\natexlab{a}})\citenamefont {Bachtis}, \citenamefont {Aarts},
  \citenamefont {Di~Renzo},\ and\ \citenamefont {Lucini}}]{Bachtis:2021eww}%
  \BibitemOpen
  \bibfield  {author} {\bibinfo {author} {\bibfnamefont {Dimitrios}\
  \bibnamefont {Bachtis}}, \bibinfo {author} {\bibfnamefont {Gert}\
  \bibnamefont {Aarts}}, \bibinfo {author} {\bibfnamefont {Francesco}\
  \bibnamefont {Di~Renzo}}, \ and\ \bibinfo {author} {\bibfnamefont {Biagio}\
  \bibnamefont {Lucini}},\ }\bibfield  {title} {\enquote {\bibinfo {title}
  {{Inverse renormalization group in quantum field theory}},}\ }\href@noop {}
  {\bibfield  {journal} {\bibinfo  {journal} {Phys. Rev. Lett. (to appear)}\ }
  (\bibinfo {year} {2021}{\natexlab{a}})},\ \Eprint
  {http://arxiv.org/abs/2107.00466} {arXiv:2107.00466 [hep-lat]} \BibitemShut
  {NoStop}%
\bibitem [{\citenamefont {Alexandru}\ \emph {et~al.}(2017)\citenamefont
  {Alexandru}, \citenamefont {Bedaque}, \citenamefont {Lamm},\ and\
  \citenamefont {Lawrence}}]{Alexandru:2017czx}%
  \BibitemOpen
  \bibfield  {author} {\bibinfo {author} {\bibfnamefont {Andrei}\ \bibnamefont
  {Alexandru}}, \bibinfo {author} {\bibfnamefont {Paulo~F.}\ \bibnamefont
  {Bedaque}}, \bibinfo {author} {\bibfnamefont {Henry}\ \bibnamefont {Lamm}}, \
  and\ \bibinfo {author} {\bibfnamefont {Scott}\ \bibnamefont {Lawrence}},\
  }\bibfield  {title} {\enquote {\bibinfo {title} {{Deep Learning Beyond
  Lefschetz Thimbles}},}\ }\href@noop {} {\bibfield  {journal} {\bibinfo
  {journal} {Phys. Rev.}\ }\textbf {\bibinfo {volume} {D96}},\ \bibinfo {pages}
  {094505} (\bibinfo {year} {2017})}\BibitemShut {NoStop}%
\bibitem [{\citenamefont {Lawrence}\ and\ \citenamefont
  {Yamauchi}(2021)}]{Lawrence:2021izu}%
  \BibitemOpen
  \bibfield  {author} {\bibinfo {author} {\bibfnamefont {Scott}\ \bibnamefont
  {Lawrence}}\ and\ \bibinfo {author} {\bibfnamefont {Yukari}\ \bibnamefont
  {Yamauchi}},\ }\bibfield  {title} {\enquote {\bibinfo {title} {{Normalizing
  Flows and the Real-Time Sign Problem}},}\ }\href {\doibase
  10.1103/PhysRevD.103.114509} {\bibfield  {journal} {\bibinfo  {journal}
  {Phys. Rev. D}\ }\textbf {\bibinfo {volume} {103}},\ \bibinfo {pages}
  {114509} (\bibinfo {year} {2021})},\ \Eprint
  {http://arxiv.org/abs/2101.05755} {arXiv:2101.05755 [hep-lat]} \BibitemShut
  {NoStop}%
\bibitem [{\citenamefont {Wynen}\ \emph {et~al.}(2021)\citenamefont {Wynen},
  \citenamefont {Berkowitz}, \citenamefont {Krieg}, \citenamefont {Luu},\ and\
  \citenamefont {Ostmeyer}}]{Wynen:2020uzx}%
  \BibitemOpen
  \bibfield  {author} {\bibinfo {author} {\bibfnamefont {Jan-Lukas}\
  \bibnamefont {Wynen}}, \bibinfo {author} {\bibfnamefont {Evan}\ \bibnamefont
  {Berkowitz}}, \bibinfo {author} {\bibfnamefont {Stefan}\ \bibnamefont
  {Krieg}}, \bibinfo {author} {\bibfnamefont {Thomas}\ \bibnamefont {Luu}}, \
  and\ \bibinfo {author} {\bibfnamefont {Johann}\ \bibnamefont {Ostmeyer}},\
  }\bibfield  {title} {\enquote {\bibinfo {title} {{Machine learning to
  alleviate Hubbard-model sign problems}},}\ }\href {\doibase
  10.1103/PhysRevB.103.125153} {\bibfield  {journal} {\bibinfo  {journal}
  {Phys. Rev. B}\ }\textbf {\bibinfo {volume} {103}},\ \bibinfo {pages}
  {125153} (\bibinfo {year} {2021})},\ \Eprint
  {http://arxiv.org/abs/2006.11221} {arXiv:2006.11221 [cond-mat.str-el]}
  \BibitemShut {NoStop}%
\bibitem [{\citenamefont {Shanahan}\ \emph {et~al.}(2018)\citenamefont
  {Shanahan}, \citenamefont {Trewartha},\ and\ \citenamefont
  {Detmold}}]{Shanahan:2018vcv}%
  \BibitemOpen
  \bibfield  {author} {\bibinfo {author} {\bibfnamefont {Phiala~E.}\
  \bibnamefont {Shanahan}}, \bibinfo {author} {\bibfnamefont {Daniel}\
  \bibnamefont {Trewartha}}, \ and\ \bibinfo {author} {\bibfnamefont {William}\
  \bibnamefont {Detmold}},\ }\bibfield  {title} {\enquote {\bibinfo {title}
  {{Machine learning action parameters in lattice quantum chromodynamics}},}\
  }\href {\doibase 10.1103/PhysRevD.97.094506} {\bibfield  {journal} {\bibinfo
  {journal} {Phys. Rev. D}\ }\textbf {\bibinfo {volume} {97}},\ \bibinfo
  {pages} {094506} (\bibinfo {year} {2018})},\ \Eprint
  {http://arxiv.org/abs/1801.05784} {arXiv:1801.05784 [hep-lat]} \BibitemShut
  {NoStop}%
\bibitem [{\citenamefont {Favoni}\ \emph {et~al.}(2022)\citenamefont {Favoni},
  \citenamefont {Ipp}, \citenamefont {M\"uller},\ and\ \citenamefont
  {Schuh}}]{Favoni:2020reg}%
  \BibitemOpen
  \bibfield  {author} {\bibinfo {author} {\bibfnamefont {Matteo}\ \bibnamefont
  {Favoni}}, \bibinfo {author} {\bibfnamefont {Andreas}\ \bibnamefont {Ipp}},
  \bibinfo {author} {\bibfnamefont {David~I.}\ \bibnamefont {M\"uller}}, \ and\
  \bibinfo {author} {\bibfnamefont {Daniel}\ \bibnamefont {Schuh}},\ }\bibfield
   {title} {\enquote {\bibinfo {title} {{Lattice Gauge Equivariant
  Convolutional Neural Networks}},}\ }\href {\doibase
  10.1103/PhysRevLett.128.032003} {\bibfield  {journal} {\bibinfo  {journal}
  {Phys. Rev. Lett.}\ }\textbf {\bibinfo {volume} {128}},\ \bibinfo {pages}
  {032003} (\bibinfo {year} {2022})},\ \Eprint
  {http://arxiv.org/abs/2012.12901} {arXiv:2012.12901 [hep-lat]} \BibitemShut
  {NoStop}%
\bibitem [{\citenamefont {Bulusu}\ \emph {et~al.}(2021)\citenamefont {Bulusu},
  \citenamefont {Favoni}, \citenamefont {Ipp}, \citenamefont {M\"uller},\ and\
  \citenamefont {Schuh}}]{Bulusu:2021rqz}%
  \BibitemOpen
  \bibfield  {author} {\bibinfo {author} {\bibfnamefont {Srinath}\ \bibnamefont
  {Bulusu}}, \bibinfo {author} {\bibfnamefont {Matteo}\ \bibnamefont {Favoni}},
  \bibinfo {author} {\bibfnamefont {Andreas}\ \bibnamefont {Ipp}}, \bibinfo
  {author} {\bibfnamefont {David~I.}\ \bibnamefont {M\"uller}}, \ and\ \bibinfo
  {author} {\bibfnamefont {Daniel}\ \bibnamefont {Schuh}},\ }\bibfield  {title}
  {\enquote {\bibinfo {title} {{Generalization capabilities of translationally
  equivariant neural networks}},}\ }\href {\doibase
  10.1103/PhysRevD.104.074504} {\bibfield  {journal} {\bibinfo  {journal}
  {Phys. Rev. D}\ }\textbf {\bibinfo {volume} {104}},\ \bibinfo {pages}
  {074504} (\bibinfo {year} {2021})},\ \Eprint
  {http://arxiv.org/abs/2103.14686} {arXiv:2103.14686 [hep-lat]} \BibitemShut
  {NoStop}%
\bibitem [{\citenamefont {Matsumoto}\ \emph {et~al.}(2021)\citenamefont
  {Matsumoto}, \citenamefont {Kitazawa},\ and\ \citenamefont
  {Kohno}}]{Matsumoto:2019jia}%
  \BibitemOpen
  \bibfield  {author} {\bibinfo {author} {\bibfnamefont {Takuya}\ \bibnamefont
  {Matsumoto}}, \bibinfo {author} {\bibfnamefont {Masakiyo}\ \bibnamefont
  {Kitazawa}}, \ and\ \bibinfo {author} {\bibfnamefont {Yasuhiro}\ \bibnamefont
  {Kohno}},\ }\bibfield  {title} {\enquote {\bibinfo {title} {{Classifying
  topological charge in SU(3) Yang\textendash{}Mills theory with machine
  learning}},}\ }\href {\doibase 10.1093/ptep/ptaa138} {\bibfield  {journal}
  {\bibinfo  {journal} {PTEP}\ }\textbf {\bibinfo {volume} {2021}},\ \bibinfo
  {pages} {023D01} (\bibinfo {year} {2021})},\ \Eprint
  {http://arxiv.org/abs/1909.06238} {arXiv:1909.06238 [hep-lat]} \BibitemShut
  {NoStop}%
\bibitem [{\citenamefont {Boyda}\ \emph
  {et~al.}(2021{\natexlab{b}})\citenamefont {Boyda}, \citenamefont {Chernodub},
  \citenamefont {Gerasimeniuk}, \citenamefont {Goy}, \citenamefont {Liubimov},\
  and\ \citenamefont {Molochkov}}]{Boyda:2020nfh}%
  \BibitemOpen
  \bibfield  {author} {\bibinfo {author} {\bibfnamefont {D.~L.}\ \bibnamefont
  {Boyda}}, \bibinfo {author} {\bibfnamefont {M.~N.}\ \bibnamefont
  {Chernodub}}, \bibinfo {author} {\bibfnamefont {N.~V.}\ \bibnamefont
  {Gerasimeniuk}}, \bibinfo {author} {\bibfnamefont {V.~A.}\ \bibnamefont
  {Goy}}, \bibinfo {author} {\bibfnamefont {S.~D.}\ \bibnamefont {Liubimov}}, \
  and\ \bibinfo {author} {\bibfnamefont {A.~V.}\ \bibnamefont {Molochkov}},\
  }\bibfield  {title} {\enquote {\bibinfo {title} {{Finding the deconfinement
  temperature in lattice Yang-Mills theories from outside the scaling window
  with machine learning}},}\ }\href {\doibase 10.1103/PhysRevD.103.014509}
  {\bibfield  {journal} {\bibinfo  {journal} {Phys. Rev. D}\ }\textbf {\bibinfo
  {volume} {103}},\ \bibinfo {pages} {014509} (\bibinfo {year}
  {2021}{\natexlab{b}})},\ \Eprint {http://arxiv.org/abs/2009.10971}
  {arXiv:2009.10971 [hep-lat]} \BibitemShut {NoStop}%
\bibitem [{\citenamefont {Bachtis}\ \emph
  {et~al.}(2021{\natexlab{b}})\citenamefont {Bachtis}, \citenamefont {Aarts},\
  and\ \citenamefont {Lucini}}]{Bachtis:2020fly}%
  \BibitemOpen
  \bibfield  {author} {\bibinfo {author} {\bibfnamefont {Dimitrios}\
  \bibnamefont {Bachtis}}, \bibinfo {author} {\bibfnamefont {Gert}\
  \bibnamefont {Aarts}}, \ and\ \bibinfo {author} {\bibfnamefont {Biagio}\
  \bibnamefont {Lucini}},\ }\bibfield  {title} {\enquote {\bibinfo {title}
  {{Adding machine learning within Hamiltonians: Renormalization group
  transformations, symmetry breaking and restoration}},}\ }\href {\doibase
  10.1103/PhysRevResearch.3.013134} {\bibfield  {journal} {\bibinfo  {journal}
  {Phys. Rev. Res.}\ }\textbf {\bibinfo {volume} {3}},\ \bibinfo {pages}
  {013134} (\bibinfo {year} {2021}{\natexlab{b}})},\ \Eprint
  {http://arxiv.org/abs/2010.00054} {arXiv:2010.00054 [hep-lat]} \BibitemShut
  {NoStop}%
\bibitem [{\citenamefont {Palermo}\ \emph {et~al.}(2021)\citenamefont
  {Palermo}, \citenamefont {Anderlini}, \citenamefont {Lombardo}, \citenamefont
  {Kotov},\ and\ \citenamefont {Trunin}}]{Palermo:2021jrf}%
  \BibitemOpen
  \bibfield  {author} {\bibinfo {author} {\bibfnamefont {Andrea}\ \bibnamefont
  {Palermo}}, \bibinfo {author} {\bibfnamefont {Lucio}\ \bibnamefont
  {Anderlini}}, \bibinfo {author} {\bibfnamefont {Maria~Paola}\ \bibnamefont
  {Lombardo}}, \bibinfo {author} {\bibfnamefont {Andrey}\ \bibnamefont
  {Kotov}}, \ and\ \bibinfo {author} {\bibfnamefont {Anton}\ \bibnamefont
  {Trunin}},\ }\bibfield  {title} {\enquote {\bibinfo {title} {{Machine
  learning approaches to the QCD transition}},}\ }in\ \href@noop {} {\emph
  {\bibinfo {booktitle} {{38th International Symposium on Lattice Field
  Theory}}}}\ (\bibinfo {year} {2021})\ \Eprint
  {http://arxiv.org/abs/2111.05216} {arXiv:2111.05216 [hep-lat]} \BibitemShut
  {NoStop}%
\bibitem [{\citenamefont {Tan}\ \emph {et~al.}(2021)\citenamefont {Tan},
  \citenamefont {Peng}, \citenamefont {Tseng},\ and\ \citenamefont
  {Jiang}}]{Tan:2021cgs}%
  \BibitemOpen
  \bibfield  {author} {\bibinfo {author} {\bibfnamefont {D.~R}\ \bibnamefont
  {Tan}}, \bibinfo {author} {\bibfnamefont {J.~H}\ \bibnamefont {Peng}},
  \bibinfo {author} {\bibfnamefont {Y.~H}\ \bibnamefont {Tseng}}, \ and\
  \bibinfo {author} {\bibfnamefont {F.~J}\ \bibnamefont {Jiang}},\ }\bibfield
  {title} {\enquote {\bibinfo {title} {{A universal neural network for learning
  phases}},}\ }\href {\doibase 10.1140/epjp/s13360-021-02121-4} {\bibfield
  {journal} {\bibinfo  {journal} {Eur. Phys. J. Plus}\ }\textbf {\bibinfo
  {volume} {136}},\ \bibinfo {pages} {1116} (\bibinfo {year} {2021})},\ \Eprint
  {http://arxiv.org/abs/2103.10846} {arXiv:2103.10846 [cond-mat.dis-nn]}
  \BibitemShut {NoStop}%
\bibitem [{\citenamefont {Li}\ \emph {et~al.}(2018)\citenamefont {Li},
  \citenamefont {Tan},\ and\ \citenamefont {Jiang}}]{Li:2017xaz}%
  \BibitemOpen
  \bibfield  {author} {\bibinfo {author} {\bibfnamefont {Chian-De}\
  \bibnamefont {Li}}, \bibinfo {author} {\bibfnamefont {Deng-Ruei}\
  \bibnamefont {Tan}}, \ and\ \bibinfo {author} {\bibfnamefont {Fu-Jiun}\
  \bibnamefont {Jiang}},\ }\bibfield  {title} {\enquote {\bibinfo {title}
  {{Applications of neural networks to the studies of phase transitions of
  two-dimensional Potts models}},}\ }\href {\doibase 10.1016/j.aop.2018.02.018}
  {\bibfield  {journal} {\bibinfo  {journal} {Annals Phys.}\ }\textbf {\bibinfo
  {volume} {391}},\ \bibinfo {pages} {312--331} (\bibinfo {year} {2018})},\
  \Eprint {http://arxiv.org/abs/1703.02369} {arXiv:1703.02369
  [cond-mat.dis-nn]} \BibitemShut {NoStop}%
\bibitem [{\citenamefont {Wetzel}\ and\ \citenamefont
  {Scherzer}(2017)}]{Wetzel:2017ooo}%
  \BibitemOpen
  \bibfield  {author} {\bibinfo {author} {\bibfnamefont {Sebastian~Johann}\
  \bibnamefont {Wetzel}}\ and\ \bibinfo {author} {\bibfnamefont {Manuel}\
  \bibnamefont {Scherzer}},\ }\bibfield  {title} {\enquote {\bibinfo {title}
  {{Machine Learning of Explicit Order Parameters: From the Ising Model to
  SU(2) Lattice Gauge Theory}},}\ }\href {\doibase 10.1103/PhysRevB.96.184410}
  {\bibfield  {journal} {\bibinfo  {journal} {Phys. Rev. B}\ }\textbf {\bibinfo
  {volume} {96}},\ \bibinfo {pages} {184410} (\bibinfo {year} {2017})},\
  \Eprint {http://arxiv.org/abs/1705.05582} {arXiv:1705.05582
  [cond-mat.stat-mech]} \BibitemShut {NoStop}%
\bibitem [{\citenamefont {Alexandrou}\ \emph {et~al.}(2020)\citenamefont
  {Alexandrou}, \citenamefont {Athenodorou}, \citenamefont {Chrysostomou},\
  and\ \citenamefont {Paul}}]{Alexandrou:2019hgt}%
  \BibitemOpen
  \bibfield  {author} {\bibinfo {author} {\bibfnamefont {Constantia}\
  \bibnamefont {Alexandrou}}, \bibinfo {author} {\bibfnamefont {Andreas}\
  \bibnamefont {Athenodorou}}, \bibinfo {author} {\bibfnamefont {Charalambos}\
  \bibnamefont {Chrysostomou}}, \ and\ \bibinfo {author} {\bibfnamefont
  {Srijit}\ \bibnamefont {Paul}},\ }\bibfield  {title} {\enquote {\bibinfo
  {title} {{The critical temperature of the 2D-Ising model through Deep
  Learning Autoencoders}},}\ }\href {\doibase 10.1140/epjb/e2020-100506-5}
  {\bibfield  {journal} {\bibinfo  {journal} {Eur. Phys. J. B}\ }\textbf
  {\bibinfo {volume} {93}},\ \bibinfo {pages} {226} (\bibinfo {year} {2020})},\
  \Eprint {http://arxiv.org/abs/1903.03506} {arXiv:1903.03506
  [cond-mat.stat-mech]} \BibitemShut {NoStop}%
\bibitem [{\citenamefont {Bl\"ucher}\ \emph {et~al.}(2020)\citenamefont
  {Bl\"ucher}, \citenamefont {Kades}, \citenamefont {Pawlowski}, \citenamefont
  {Strodthoff},\ and\ \citenamefont {Urban}}]{Blucher:2020mjt}%
  \BibitemOpen
  \bibfield  {author} {\bibinfo {author} {\bibfnamefont {Stefan}\ \bibnamefont
  {Bl\"ucher}}, \bibinfo {author} {\bibfnamefont {Lukas}\ \bibnamefont
  {Kades}}, \bibinfo {author} {\bibfnamefont {Jan~M.}\ \bibnamefont
  {Pawlowski}}, \bibinfo {author} {\bibfnamefont {Nils}\ \bibnamefont
  {Strodthoff}}, \ and\ \bibinfo {author} {\bibfnamefont {Julian~M.}\
  \bibnamefont {Urban}},\ }\bibfield  {title} {\enquote {\bibinfo {title}
  {{Towards novel insights in lattice field theory with explainable machine
  learning}},}\ }\href {\doibase 10.1103/PhysRevD.101.094507} {\bibfield
  {journal} {\bibinfo  {journal} {Phys. Rev. D}\ }\textbf {\bibinfo {volume}
  {101}},\ \bibinfo {pages} {094507} (\bibinfo {year} {2020})},\ \Eprint
  {http://arxiv.org/abs/2003.01504} {arXiv:2003.01504 [hep-lat]} \BibitemShut
  {NoStop}%
\bibitem [{\citenamefont {Yau}\ and\ \citenamefont {Su}(2020)}]{Yau:2020emg}%
  \BibitemOpen
  \bibfield  {author} {\bibinfo {author} {\bibfnamefont {Hon~Man}\ \bibnamefont
  {Yau}}\ and\ \bibinfo {author} {\bibfnamefont {Nan}\ \bibnamefont {Su}},\
  }\bibfield  {title} {\enquote {\bibinfo {title} {{On the generalizability of
  artificial neural networks in spin models}},}\ }\href@noop {} {\  (\bibinfo
  {year} {2020})},\ \Eprint {http://arxiv.org/abs/2006.15021} {arXiv:2006.15021
  [cond-mat.dis-nn]} \BibitemShut {NoStop}%
\bibitem [{\citenamefont {Zhou}\ \emph {et~al.}(2019)\citenamefont {Zhou},
  \citenamefont {Endr\H{o}di}, \citenamefont {Pang},\ and\ \citenamefont
  {St\"ocker}}]{Zhou:2018ill}%
  \BibitemOpen
  \bibfield  {author} {\bibinfo {author} {\bibfnamefont {Kai}\ \bibnamefont
  {Zhou}}, \bibinfo {author} {\bibfnamefont {Gergely}\ \bibnamefont
  {Endr\H{o}di}}, \bibinfo {author} {\bibfnamefont {Long-Gang}\ \bibnamefont
  {Pang}}, \ and\ \bibinfo {author} {\bibfnamefont {Horst}\ \bibnamefont
  {St\"ocker}},\ }\bibfield  {title} {\enquote {\bibinfo {title} {{Regressive
  and generative neural networks for scalar field theory}},}\ }\href {\doibase
  10.1103/PhysRevD.100.011501} {\bibfield  {journal} {\bibinfo  {journal}
  {Phys. Rev. D}\ }\textbf {\bibinfo {volume} {100}},\ \bibinfo {pages}
  {011501} (\bibinfo {year} {2019})},\ \Eprint
  {http://arxiv.org/abs/1810.12879} {arXiv:1810.12879 [hep-lat]} \BibitemShut
  {NoStop}%
\bibitem [{\citenamefont {Detmold}\ \emph {et~al.}(2021)\citenamefont
  {Detmold}, \citenamefont {Kanwar}, \citenamefont {Lamm}, \citenamefont
  {Wagman},\ and\ \citenamefont {Warrington}}]{Detmold:2021ulb}%
  \BibitemOpen
  \bibfield  {author} {\bibinfo {author} {\bibfnamefont {William}\ \bibnamefont
  {Detmold}}, \bibinfo {author} {\bibfnamefont {Gurtej}\ \bibnamefont
  {Kanwar}}, \bibinfo {author} {\bibfnamefont {Henry}\ \bibnamefont {Lamm}},
  \bibinfo {author} {\bibfnamefont {Michael~L.}\ \bibnamefont {Wagman}}, \ and\
  \bibinfo {author} {\bibfnamefont {Neill~C.}\ \bibnamefont {Warrington}},\
  }\bibfield  {title} {\enquote {\bibinfo {title} {{Path integral contour
  deformations for observables in $SU(N)$ gauge theory}},}\ }\href {\doibase
  10.1103/PhysRevD.103.094517} {\bibfield  {journal} {\bibinfo  {journal}
  {Phys. Rev. D}\ }\textbf {\bibinfo {volume} {103}},\ \bibinfo {pages}
  {094517} (\bibinfo {year} {2021})},\ \Eprint
  {http://arxiv.org/abs/2101.12668} {arXiv:2101.12668 [hep-lat]} \BibitemShut
  {NoStop}%
\bibitem [{\citenamefont {Yoon}\ \emph {et~al.}(2019)\citenamefont {Yoon},
  \citenamefont {Bhattacharya},\ and\ \citenamefont {Gupta}}]{Yoon:2018krb}%
  \BibitemOpen
  \bibfield  {author} {\bibinfo {author} {\bibfnamefont {Boram}\ \bibnamefont
  {Yoon}}, \bibinfo {author} {\bibfnamefont {Tanmoy}\ \bibnamefont
  {Bhattacharya}}, \ and\ \bibinfo {author} {\bibfnamefont {Rajan}\
  \bibnamefont {Gupta}},\ }\bibfield  {title} {\enquote {\bibinfo {title}
  {{Machine Learning Estimators for Lattice QCD Observables}},}\ }\href
  {\doibase 10.1103/PhysRevD.100.014504} {\bibfield  {journal} {\bibinfo
  {journal} {Phys. Rev. D}\ }\textbf {\bibinfo {volume} {100}},\ \bibinfo
  {pages} {014504} (\bibinfo {year} {2019})},\ \Eprint
  {http://arxiv.org/abs/1807.05971} {arXiv:1807.05971 [hep-lat]} \BibitemShut
  {NoStop}%
\bibitem [{\citenamefont {Zhang}\ \emph {et~al.}(2020)\citenamefont {Zhang},
  \citenamefont {Fan}, \citenamefont {Li}, \citenamefont {Lin},\ and\
  \citenamefont {Yoon}}]{Zhang:2019qiq}%
  \BibitemOpen
  \bibfield  {author} {\bibinfo {author} {\bibfnamefont {Rui}\ \bibnamefont
  {Zhang}}, \bibinfo {author} {\bibfnamefont {Zhouyou}\ \bibnamefont {Fan}},
  \bibinfo {author} {\bibfnamefont {Ruizi}\ \bibnamefont {Li}}, \bibinfo
  {author} {\bibfnamefont {Huey-Wen}\ \bibnamefont {Lin}}, \ and\ \bibinfo
  {author} {\bibfnamefont {Boram}\ \bibnamefont {Yoon}},\ }\bibfield  {title}
  {\enquote {\bibinfo {title} {{Machine-learning prediction for quasiparton
  distribution function matrix elements}},}\ }\href {\doibase
  10.1103/PhysRevD.101.034516} {\bibfield  {journal} {\bibinfo  {journal}
  {Phys. Rev. D}\ }\textbf {\bibinfo {volume} {101}},\ \bibinfo {pages}
  {034516} (\bibinfo {year} {2020})},\ \Eprint
  {http://arxiv.org/abs/1909.10990} {arXiv:1909.10990 [hep-lat]} \BibitemShut
  {NoStop}%
\bibitem [{\citenamefont {Hudspith}\ and\ \citenamefont
  {Mohler}(2021)}]{Hudspith:2021iqu}%
  \BibitemOpen
  \bibfield  {author} {\bibinfo {author} {\bibfnamefont {R.~J.}\ \bibnamefont
  {Hudspith}}\ and\ \bibinfo {author} {\bibfnamefont {D.}~\bibnamefont
  {Mohler}},\ }\bibfield  {title} {\enquote {\bibinfo {title} {{A fully
  non-perturbative charm-quark tuning using machine learning}},}\ }\href@noop
  {} {\  (\bibinfo {year} {2021})},\ \Eprint {http://arxiv.org/abs/2112.01997}
  {arXiv:2112.01997 [hep-lat]} \BibitemShut {NoStop}%
\bibitem [{\citenamefont {Kades}\ \emph {et~al.}(2020)\citenamefont {Kades},
  \citenamefont {Pawlowski}, \citenamefont {Rothkopf}, \citenamefont
  {Scherzer}, \citenamefont {Urban}, \citenamefont {Wetzel}, \citenamefont
  {Wink},\ and\ \citenamefont {Ziegler}}]{Kades:2019wtd}%
  \BibitemOpen
  \bibfield  {author} {\bibinfo {author} {\bibfnamefont {Lukas}\ \bibnamefont
  {Kades}}, \bibinfo {author} {\bibfnamefont {Jan~M.}\ \bibnamefont
  {Pawlowski}}, \bibinfo {author} {\bibfnamefont {Alexander}\ \bibnamefont
  {Rothkopf}}, \bibinfo {author} {\bibfnamefont {Manuel}\ \bibnamefont
  {Scherzer}}, \bibinfo {author} {\bibfnamefont {Julian~M.}\ \bibnamefont
  {Urban}}, \bibinfo {author} {\bibfnamefont {Sebastian~J.}\ \bibnamefont
  {Wetzel}}, \bibinfo {author} {\bibfnamefont {Nicolas}\ \bibnamefont {Wink}},
  \ and\ \bibinfo {author} {\bibfnamefont {Felix P.~G.}\ \bibnamefont
  {Ziegler}},\ }\bibfield  {title} {\enquote {\bibinfo {title} {{Spectral
  Reconstruction with Deep Neural Networks}},}\ }\href {\doibase
  10.1103/PhysRevD.102.096001} {\bibfield  {journal} {\bibinfo  {journal}
  {Phys. Rev. D}\ }\textbf {\bibinfo {volume} {102}},\ \bibinfo {pages}
  {096001} (\bibinfo {year} {2020})},\ \Eprint
  {http://arxiv.org/abs/1905.04305} {arXiv:1905.04305 [physics.comp-ph]}
  \BibitemShut {NoStop}%
\bibitem [{\citenamefont {Offler}\ \emph {et~al.}(2019)\citenamefont {Offler},
  \citenamefont {Aarts}, \citenamefont {Allton}, \citenamefont {Glesaaen},
  \citenamefont {J\"ager}, \citenamefont {Kim}, \citenamefont {Lombardo},
  \citenamefont {Ryan},\ and\ \citenamefont {Skullerud}}]{Offler:2019eij}%
  \BibitemOpen
  \bibfield  {author} {\bibinfo {author} {\bibfnamefont {Sam}\ \bibnamefont
  {Offler}}, \bibinfo {author} {\bibfnamefont {Gert}\ \bibnamefont {Aarts}},
  \bibinfo {author} {\bibfnamefont {Chris}\ \bibnamefont {Allton}}, \bibinfo
  {author} {\bibfnamefont {Jonas}\ \bibnamefont {Glesaaen}}, \bibinfo {author}
  {\bibfnamefont {Benjamin}\ \bibnamefont {J\"ager}}, \bibinfo {author}
  {\bibfnamefont {Seyong}\ \bibnamefont {Kim}}, \bibinfo {author}
  {\bibfnamefont {Maria~Paola}\ \bibnamefont {Lombardo}}, \bibinfo {author}
  {\bibfnamefont {Sinead~M.}\ \bibnamefont {Ryan}}, \ and\ \bibinfo {author}
  {\bibfnamefont {Jon-Ivar}\ \bibnamefont {Skullerud}},\ }\bibfield  {title}
  {\enquote {\bibinfo {title} {{News from bottomonium spectral functions in
  thermal QCD}},}\ }\href {\doibase 10.22323/1.363.0076} {\bibfield  {journal}
  {\bibinfo  {journal} {PoS}\ }\textbf {\bibinfo {volume} {LATTICE2019}},\
  \bibinfo {pages} {076} (\bibinfo {year} {2019})},\ \Eprint
  {http://arxiv.org/abs/1912.12900} {arXiv:1912.12900 [hep-lat]} \BibitemShut
  {NoStop}%
\bibitem [{\citenamefont {Horak}\ \emph {et~al.}(2021)\citenamefont {Horak},
  \citenamefont {Pawlowski}, \citenamefont {Rodr\'\i{}guez-Quintero},
  \citenamefont {Turnwald}, \citenamefont {Urban}, \citenamefont {Wink},\ and\
  \citenamefont {Zafeiropoulos}}]{Horak:2021syv}%
  \BibitemOpen
  \bibfield  {author} {\bibinfo {author} {\bibfnamefont {Jan}\ \bibnamefont
  {Horak}}, \bibinfo {author} {\bibfnamefont {Jan~M.}\ \bibnamefont
  {Pawlowski}}, \bibinfo {author} {\bibfnamefont {Jos\'e}\ \bibnamefont
  {Rodr\'\i{}guez-Quintero}}, \bibinfo {author} {\bibfnamefont {Jonas}\
  \bibnamefont {Turnwald}}, \bibinfo {author} {\bibfnamefont {Julian~M.}\
  \bibnamefont {Urban}}, \bibinfo {author} {\bibfnamefont {Nicolas}\
  \bibnamefont {Wink}}, \ and\ \bibinfo {author} {\bibfnamefont {Savvas}\
  \bibnamefont {Zafeiropoulos}},\ }\bibfield  {title} {\enquote {\bibinfo
  {title} {{Reconstructing QCD Spectral Functions with Gaussian Processes}},}\
  }\href@noop {} {\  (\bibinfo {year} {2021})},\ \Eprint
  {http://arxiv.org/abs/2107.13464} {arXiv:2107.13464 [hep-ph]} \BibitemShut
  {NoStop}%
\bibitem [{\citenamefont {Chen}\ \emph {et~al.}(2021)\citenamefont {Chen},
  \citenamefont {Ding}, \citenamefont {Liu}, \citenamefont {Papp},\ and\
  \citenamefont {Yang}}]{Chen:2021giw}%
  \BibitemOpen
  \bibfield  {author} {\bibinfo {author} {\bibfnamefont {S.~Y.}\ \bibnamefont
  {Chen}}, \bibinfo {author} {\bibfnamefont {H.~T.}\ \bibnamefont {Ding}},
  \bibinfo {author} {\bibfnamefont {F.~Y.}\ \bibnamefont {Liu}}, \bibinfo
  {author} {\bibfnamefont {G.}~\bibnamefont {Papp}}, \ and\ \bibinfo {author}
  {\bibfnamefont {C.~B.}\ \bibnamefont {Yang}},\ }\bibfield  {title} {\enquote
  {\bibinfo {title} {{Machine learning spectral functions in lattice QCD}},}\
  }\href@noop {} {\  (\bibinfo {year} {2021})},\ \Eprint
  {http://arxiv.org/abs/2110.13521} {arXiv:2110.13521 [hep-lat]} \BibitemShut
  {NoStop}%
\bibitem [{\citenamefont {Wang}\ \emph {et~al.}(2021)\citenamefont {Wang},
  \citenamefont {Shi},\ and\ \citenamefont {Zhou}}]{Wang:2021cqw}%
  \BibitemOpen
  \bibfield  {author} {\bibinfo {author} {\bibfnamefont {Lingxiao}\
  \bibnamefont {Wang}}, \bibinfo {author} {\bibfnamefont {Shuzhe}\ \bibnamefont
  {Shi}}, \ and\ \bibinfo {author} {\bibfnamefont {Kai}\ \bibnamefont {Zhou}},\
  }\bibfield  {title} {\enquote {\bibinfo {title} {{Automatic differentiation
  approach for reconstructing spectral functions with neural networks}},}\ }in\
  \href@noop {} {\emph {\bibinfo {booktitle} {{35th Conference on Neural
  Information Processing Systems}}}}\ (\bibinfo {year} {2021})\ \Eprint
  {http://arxiv.org/abs/2112.06206} {arXiv:2112.06206 [physics.comp-ph]}
  \BibitemShut {NoStop}%
\bibitem [{\citenamefont {Shi}\ \emph {et~al.}(2022)\citenamefont {Shi},
  \citenamefont {Wang},\ and\ \citenamefont {Zhou}}]{Shi:2022yqw}%
  \BibitemOpen
  \bibfield  {author} {\bibinfo {author} {\bibfnamefont {Shuzhe}\ \bibnamefont
  {Shi}}, \bibinfo {author} {\bibfnamefont {Lingxiao}\ \bibnamefont {Wang}}, \
  and\ \bibinfo {author} {\bibfnamefont {Kai}\ \bibnamefont {Zhou}},\
  }\bibfield  {title} {\enquote {\bibinfo {title} {{Rethinking the
  ill-posedness of the spectral function reconstruction - why is it
  fundamentally hard and how Artificial Neural Networks can help}},}\
  }\href@noop {} {\  (\bibinfo {year} {2022})},\ \Eprint
  {http://arxiv.org/abs/2201.02564} {arXiv:2201.02564 [hep-ph]} \BibitemShut
  {NoStop}%
\bibitem [{\citenamefont {Duane}\ \emph {et~al.}(1987)\citenamefont {Duane},
  \citenamefont {Kennedy}, \citenamefont {Pendleton},\ and\ \citenamefont
  {Roweth}}]{Duane:1987de}%
  \BibitemOpen
  \bibfield  {author} {\bibinfo {author} {\bibfnamefont {S.}~\bibnamefont
  {Duane}}, \bibinfo {author} {\bibfnamefont {A.~D.}\ \bibnamefont {Kennedy}},
  \bibinfo {author} {\bibfnamefont {B.~J.}\ \bibnamefont {Pendleton}}, \ and\
  \bibinfo {author} {\bibfnamefont {D.}~\bibnamefont {Roweth}},\ }\bibfield
  {title} {\enquote {\bibinfo {title} {{Hybrid Monte Carlo}},}\ }\href
  {\doibase 10.1016/0370-2693(87)91197-X} {\bibfield  {journal} {\bibinfo
  {journal} {Phys. Lett. B}\ }\textbf {\bibinfo {volume} {195}},\ \bibinfo
  {pages} {216--222} (\bibinfo {year} {1987})}\BibitemShut {NoStop}%
\bibitem [{\citenamefont {Bachtis}\ \emph
  {et~al.}(2021{\natexlab{c}})\citenamefont {Bachtis}, \citenamefont {Aarts},\
  and\ \citenamefont {Lucini}}]{Bachtis:2021xoh}%
  \BibitemOpen
  \bibfield  {author} {\bibinfo {author} {\bibfnamefont {Dimitrios}\
  \bibnamefont {Bachtis}}, \bibinfo {author} {\bibfnamefont {Gert}\
  \bibnamefont {Aarts}}, \ and\ \bibinfo {author} {\bibfnamefont {Biagio}\
  \bibnamefont {Lucini}},\ }\bibfield  {title} {\enquote {\bibinfo {title}
  {{Quantum field-theoretic machine learning}},}\ }\href {\doibase
  10.1103/PhysRevD.103.074510} {\bibfield  {journal} {\bibinfo  {journal}
  {Phys. Rev. D}\ }\textbf {\bibinfo {volume} {103}},\ \bibinfo {pages}
  {074510} (\bibinfo {year} {2021}{\natexlab{c}})},\ \Eprint
  {http://arxiv.org/abs/2102.09449} {arXiv:2102.09449 [hep-lat]} \BibitemShut
  {NoStop}%
\bibitem [{\citenamefont {Clark}\ \emph {et~al.}(2010)\citenamefont {Clark},
  \citenamefont {Babich}, \citenamefont {Barros}, \citenamefont {Brower},\ and\
  \citenamefont {Rebbi}}]{Clark:2009wm}%
  \BibitemOpen
  \bibfield  {author} {\bibinfo {author} {\bibfnamefont {M.~A.}\ \bibnamefont
  {Clark}}, \bibinfo {author} {\bibfnamefont {R.}~\bibnamefont {Babich}},
  \bibinfo {author} {\bibfnamefont {K.}~\bibnamefont {Barros}}, \bibinfo
  {author} {\bibfnamefont {R.~C.}\ \bibnamefont {Brower}}, \ and\ \bibinfo
  {author} {\bibfnamefont {C.}~\bibnamefont {Rebbi}},\ }\bibfield  {title}
  {\enquote {\bibinfo {title} {{Solving Lattice QCD systems of equations using
  mixed precision solvers on GPUs}},}\ }\href {\doibase
  10.1016/j.cpc.2010.05.002} {\bibfield  {journal} {\bibinfo  {journal}
  {Comput. Phys. Commun.}\ }\textbf {\bibinfo {volume} {181}},\ \bibinfo
  {pages} {1517--1528} (\bibinfo {year} {2010})},\ \Eprint
  {http://arxiv.org/abs/0911.3191} {arXiv:0911.3191 [hep-lat]} \BibitemShut
  {NoStop}%
\bibitem [{\citenamefont {Tu}\ \emph {et~al.}(2021)\citenamefont {Tu},
  \citenamefont {Clark}, \citenamefont {Jung},\ and\ \citenamefont
  {Mawhinney}}]{Tu:2021dvv}%
  \BibitemOpen
  \bibfield  {author} {\bibinfo {author} {\bibfnamefont {Jiqun}\ \bibnamefont
  {Tu}}, \bibinfo {author} {\bibfnamefont {M.~A.}\ \bibnamefont {Clark}},
  \bibinfo {author} {\bibfnamefont {Chulwoo}\ \bibnamefont {Jung}}, \ and\
  \bibinfo {author} {\bibfnamefont {Robert}\ \bibnamefont {Mawhinney}},\
  }\bibfield  {title} {\enquote {\bibinfo {title} {{Solving DWF Dirac Equation
  Using Multi-splitting Preconditioned Conjugate Gradient with Tensor Cores on
  NVIDIA GPUs}},}\ }\href {\doibase 10.1145/3468267.3470613} {\  (\bibinfo
  {year} {2021}),\ 10.1145/3468267.3470613},\ \Eprint
  {http://arxiv.org/abs/2104.05615} {arXiv:2104.05615 [hep-lat]} \BibitemShut
  {NoStop}%
\bibitem [{in_({\natexlab{a}})}]{in_prep_symm_wp}%
  \BibitemOpen
  \enquote {\bibinfo {title} {{Symmetry Group Equivariant Architectures for
  Physics}},}\ in\ \href@noop {} {\emph {\bibinfo {booktitle} {{Snowmass
  2022}}}},\ \bibinfo {note} {{in preparation}}\BibitemShut {NoStop}%
\bibitem [{\citenamefont {Boyle}\ \emph {et~al.}(2015)\citenamefont {Boyle},
  \citenamefont {Yamaguchi}, \citenamefont {Cossu},\ and\ \citenamefont
  {Portelli}}]{Boyle:2015tjk}%
  \BibitemOpen
  \bibfield  {author} {\bibinfo {author} {\bibfnamefont {Peter}\ \bibnamefont
  {Boyle}}, \bibinfo {author} {\bibfnamefont {Azusa}\ \bibnamefont
  {Yamaguchi}}, \bibinfo {author} {\bibfnamefont {Guido}\ \bibnamefont
  {Cossu}}, \ and\ \bibinfo {author} {\bibfnamefont {Antonin}\ \bibnamefont
  {Portelli}},\ }\href@noop {} {\enquote {\bibinfo {title} {{Grid: A next
  generation data parallel C++ QCD library}},}\ } (\bibinfo {year} {2015}),\
  \Eprint {http://arxiv.org/abs/1512.03487} {arXiv:1512.03487 [hep-lat]}
  \BibitemShut {NoStop}%
\bibitem [{\citenamefont {L{\"u}scher}\ \emph {et~al.}(2022)\citenamefont
  {L{\"u}scher} \emph {et~al.}}]{openqcd}%
  \BibitemOpen
  \bibfield  {author} {\bibinfo {author} {\bibfnamefont {M.}~\bibnamefont
  {L{\"u}scher}} \emph {et~al.},\ }\href@noop {} {\enquote {\bibinfo {title}
  {{openQCD}},}\ }\bibinfo {howpublished}
  {\url{https://luscher.web.cern.ch/luscher/openQCD/}} (\bibinfo {year}
  {2022})\BibitemShut {NoStop}%
\bibitem [{\citenamefont {Edwards}\ \emph {et~al.}(2022)\citenamefont {Edwards}
  \emph {et~al.}}]{scidac}%
  \BibitemOpen
  \bibfield  {author} {\bibinfo {author} {\bibfnamefont {R.}~\bibnamefont
  {Edwards}} \emph {et~al.} (\bibinfo {collaboration} {USQCD}),\ }\href@noop {}
  {\enquote {\bibinfo {title} {{SciDAC}},}\ }\bibinfo {howpublished}
  {\url{https://usqcd-software.github.io/}} (\bibinfo {year}
  {2022})\BibitemShut {NoStop}%
\bibitem [{\citenamefont {Paszke}\ \emph {et~al.}(2019)\citenamefont {Paszke},
  \citenamefont {Gross}, \citenamefont {Massa}, \citenamefont {Lerer},
  \citenamefont {Bradbury}, \citenamefont {Chanan}, \citenamefont {Killeen},
  \citenamefont {Lin}, \citenamefont {Gimelshein}, \citenamefont {Antiga},
  \citenamefont {Desmaison}, \citenamefont {Kopf}, \citenamefont {Yang},
  \citenamefont {DeVito}, \citenamefont {Raison}, \citenamefont {Tejani},
  \citenamefont {Chilamkurthy}, \citenamefont {Steiner}, \citenamefont {Fang},
  \citenamefont {Bai},\ and\ \citenamefont {Chintala}}]{NEURIPS2019_9015}%
  \BibitemOpen
  \bibfield  {author} {\bibinfo {author} {\bibfnamefont {Adam}\ \bibnamefont
  {Paszke}}, \bibinfo {author} {\bibfnamefont {Sam}\ \bibnamefont {Gross}},
  \bibinfo {author} {\bibfnamefont {Francisco}\ \bibnamefont {Massa}}, \bibinfo
  {author} {\bibfnamefont {Adam}\ \bibnamefont {Lerer}}, \bibinfo {author}
  {\bibfnamefont {James}\ \bibnamefont {Bradbury}}, \bibinfo {author}
  {\bibfnamefont {Gregory}\ \bibnamefont {Chanan}}, \bibinfo {author}
  {\bibfnamefont {Trevor}\ \bibnamefont {Killeen}}, \bibinfo {author}
  {\bibfnamefont {Zeming}\ \bibnamefont {Lin}}, \bibinfo {author}
  {\bibfnamefont {Natalia}\ \bibnamefont {Gimelshein}}, \bibinfo {author}
  {\bibfnamefont {Luca}\ \bibnamefont {Antiga}}, \bibinfo {author}
  {\bibfnamefont {Alban}\ \bibnamefont {Desmaison}}, \bibinfo {author}
  {\bibfnamefont {Andreas}\ \bibnamefont {Kopf}}, \bibinfo {author}
  {\bibfnamefont {Edward}\ \bibnamefont {Yang}}, \bibinfo {author}
  {\bibfnamefont {Zachary}\ \bibnamefont {DeVito}}, \bibinfo {author}
  {\bibfnamefont {Martin}\ \bibnamefont {Raison}}, \bibinfo {author}
  {\bibfnamefont {Alykhan}\ \bibnamefont {Tejani}}, \bibinfo {author}
  {\bibfnamefont {Sasank}\ \bibnamefont {Chilamkurthy}}, \bibinfo {author}
  {\bibfnamefont {Benoit}\ \bibnamefont {Steiner}}, \bibinfo {author}
  {\bibfnamefont {Lu}~\bibnamefont {Fang}}, \bibinfo {author} {\bibfnamefont
  {Junjie}\ \bibnamefont {Bai}}, \ and\ \bibinfo {author} {\bibfnamefont
  {Soumith}\ \bibnamefont {Chintala}},\ }\bibfield  {title} {\enquote {\bibinfo
  {title} {Pytorch: An imperative style, high-performance deep learning
  library},}\ }in\ \href
  {http://papers.neurips.cc/paper/9015-pytorch-an-imperative-style-high-performance-deep-learning-library.pdf}
  {\emph {\bibinfo {booktitle} {Advances in Neural Information Processing
  Systems 32}}},\ \bibinfo {editor} {edited by\ \bibinfo {editor}
  {\bibfnamefont {H.}~\bibnamefont {Wallach}}, \bibinfo {editor} {\bibfnamefont
  {H.}~\bibnamefont {Larochelle}}, \bibinfo {editor} {\bibfnamefont
  {A.}~\bibnamefont {Beygelzimer}}, \bibinfo {editor} {\bibfnamefont
  {F.}~\bibnamefont {d\textquotesingle Alch\'{e}-Buc}}, \bibinfo {editor}
  {\bibfnamefont {E.}~\bibnamefont {Fox}}, \ and\ \bibinfo {editor}
  {\bibfnamefont {R.}~\bibnamefont {Garnett}}}\ (\bibinfo  {publisher} {Curran
  Associates, Inc.},\ \bibinfo {year} {2019})\ pp.\ \bibinfo {pages}
  {8024--8035}\BibitemShut {NoStop}%
\bibitem [{\citenamefont {Abadi}\ \emph {et~al.}(2015)\citenamefont {Abadi},
  \citenamefont {Agarwal}, \citenamefont {Barham}, \citenamefont {Brevdo},
  \citenamefont {Chen}, \citenamefont {Citro}, \citenamefont {Corrado},
  \citenamefont {Davis}, \citenamefont {Dean}, \citenamefont {Devin},
  \citenamefont {Ghemawat}, \citenamefont {Goodfellow}, \citenamefont {Harp},
  \citenamefont {Irving}, \citenamefont {Isard}, \citenamefont {Jia},
  \citenamefont {Jozefowicz}, \citenamefont {Kaiser}, \citenamefont {Kudlur},
  \citenamefont {Levenberg}, \citenamefont {Man\'{e}}, \citenamefont {Monga},
  \citenamefont {Moore}, \citenamefont {Murray}, \citenamefont {Olah},
  \citenamefont {Schuster}, \citenamefont {Shlens}, \citenamefont {Steiner},
  \citenamefont {Sutskever}, \citenamefont {Talwar}, \citenamefont {Tucker},
  \citenamefont {Vanhoucke}, \citenamefont {Vasudevan}, \citenamefont
  {Vi\'{e}gas}, \citenamefont {Vinyals}, \citenamefont {Warden}, \citenamefont
  {Wattenberg}, \citenamefont {Wicke}, \citenamefont {Yu},\ and\ \citenamefont
  {Zheng}}]{tensorflow2015-whitepaper}%
  \BibitemOpen
  \bibfield  {author} {\bibinfo {author} {\bibfnamefont {Mart\'{i}n}\
  \bibnamefont {Abadi}}, \bibinfo {author} {\bibfnamefont {Ashish}\
  \bibnamefont {Agarwal}}, \bibinfo {author} {\bibfnamefont {Paul}\
  \bibnamefont {Barham}}, \bibinfo {author} {\bibfnamefont {Eugene}\
  \bibnamefont {Brevdo}}, \bibinfo {author} {\bibfnamefont {Zhifeng}\
  \bibnamefont {Chen}}, \bibinfo {author} {\bibfnamefont {Craig}\ \bibnamefont
  {Citro}}, \bibinfo {author} {\bibfnamefont {Greg~S.}\ \bibnamefont
  {Corrado}}, \bibinfo {author} {\bibfnamefont {Andy}\ \bibnamefont {Davis}},
  \bibinfo {author} {\bibfnamefont {Jeffrey}\ \bibnamefont {Dean}}, \bibinfo
  {author} {\bibfnamefont {Matthieu}\ \bibnamefont {Devin}}, \bibinfo {author}
  {\bibfnamefont {Sanjay}\ \bibnamefont {Ghemawat}}, \bibinfo {author}
  {\bibfnamefont {Ian}\ \bibnamefont {Goodfellow}}, \bibinfo {author}
  {\bibfnamefont {Andrew}\ \bibnamefont {Harp}}, \bibinfo {author}
  {\bibfnamefont {Geoffrey}\ \bibnamefont {Irving}}, \bibinfo {author}
  {\bibfnamefont {Michael}\ \bibnamefont {Isard}}, \bibinfo {author}
  {\bibfnamefont {Yangqing}\ \bibnamefont {Jia}}, \bibinfo {author}
  {\bibfnamefont {Rafal}\ \bibnamefont {Jozefowicz}}, \bibinfo {author}
  {\bibfnamefont {Lukasz}\ \bibnamefont {Kaiser}}, \bibinfo {author}
  {\bibfnamefont {Manjunath}\ \bibnamefont {Kudlur}}, \bibinfo {author}
  {\bibfnamefont {Josh}\ \bibnamefont {Levenberg}}, \bibinfo {author}
  {\bibfnamefont {Dandelion}\ \bibnamefont {Man\'{e}}}, \bibinfo {author}
  {\bibfnamefont {Rajat}\ \bibnamefont {Monga}}, \bibinfo {author}
  {\bibfnamefont {Sherry}\ \bibnamefont {Moore}}, \bibinfo {author}
  {\bibfnamefont {Derek}\ \bibnamefont {Murray}}, \bibinfo {author}
  {\bibfnamefont {Chris}\ \bibnamefont {Olah}}, \bibinfo {author}
  {\bibfnamefont {Mike}\ \bibnamefont {Schuster}}, \bibinfo {author}
  {\bibfnamefont {Jonathon}\ \bibnamefont {Shlens}}, \bibinfo {author}
  {\bibfnamefont {Benoit}\ \bibnamefont {Steiner}}, \bibinfo {author}
  {\bibfnamefont {Ilya}\ \bibnamefont {Sutskever}}, \bibinfo {author}
  {\bibfnamefont {Kunal}\ \bibnamefont {Talwar}}, \bibinfo {author}
  {\bibfnamefont {Paul}\ \bibnamefont {Tucker}}, \bibinfo {author}
  {\bibfnamefont {Vincent}\ \bibnamefont {Vanhoucke}}, \bibinfo {author}
  {\bibfnamefont {Vijay}\ \bibnamefont {Vasudevan}}, \bibinfo {author}
  {\bibfnamefont {Fernanda}\ \bibnamefont {Vi\'{e}gas}}, \bibinfo {author}
  {\bibfnamefont {Oriol}\ \bibnamefont {Vinyals}}, \bibinfo {author}
  {\bibfnamefont {Pete}\ \bibnamefont {Warden}}, \bibinfo {author}
  {\bibfnamefont {Martin}\ \bibnamefont {Wattenberg}}, \bibinfo {author}
  {\bibfnamefont {Martin}\ \bibnamefont {Wicke}}, \bibinfo {author}
  {\bibfnamefont {Yuan}\ \bibnamefont {Yu}}, \ and\ \bibinfo {author}
  {\bibfnamefont {Xiaoqiang}\ \bibnamefont {Zheng}},\ }\href
  {https://www.tensorflow.org/} {\enquote {\bibinfo {title} {{TensorFlow}:
  Large-scale machine learning on heterogeneous systems},}\ } (\bibinfo {year}
  {2015}),\ \bibinfo {note} {software available from
  tensorflow.org}\BibitemShut {NoStop}%
\bibitem [{\citenamefont {Bradbury}\ \emph {et~al.}(2018)\citenamefont
  {Bradbury}, \citenamefont {Frostig}, \citenamefont {Hawkins}, \citenamefont
  {Johnson}, \citenamefont {Leary}, \citenamefont {Maclaurin}, \citenamefont
  {Necula}, \citenamefont {Paszke}, \citenamefont {Vander{P}las}, \citenamefont
  {Wanderman-{M}ilne},\ and\ \citenamefont {Zhang}}]{jax2018github}%
  \BibitemOpen
  \bibfield  {author} {\bibinfo {author} {\bibfnamefont {James}\ \bibnamefont
  {Bradbury}}, \bibinfo {author} {\bibfnamefont {Roy}\ \bibnamefont {Frostig}},
  \bibinfo {author} {\bibfnamefont {Peter}\ \bibnamefont {Hawkins}}, \bibinfo
  {author} {\bibfnamefont {Matthew~James}\ \bibnamefont {Johnson}}, \bibinfo
  {author} {\bibfnamefont {Chris}\ \bibnamefont {Leary}}, \bibinfo {author}
  {\bibfnamefont {Dougal}\ \bibnamefont {Maclaurin}}, \bibinfo {author}
  {\bibfnamefont {George}\ \bibnamefont {Necula}}, \bibinfo {author}
  {\bibfnamefont {Adam}\ \bibnamefont {Paszke}}, \bibinfo {author}
  {\bibfnamefont {Jake}\ \bibnamefont {Vander{P}las}}, \bibinfo {author}
  {\bibfnamefont {Skye}\ \bibnamefont {Wanderman-{M}ilne}}, \ and\ \bibinfo
  {author} {\bibfnamefont {Qiao}\ \bibnamefont {Zhang}},\ }\href
  {http://github.com/google/jax} {\enquote {\bibinfo {title} {{JAX}: composable
  transformations of {P}ython+{N}um{P}y programs},}\ } (\bibinfo {year}
  {2018})\BibitemShut {NoStop}%
\bibitem [{\citenamefont {Sergeev}\ and\ \citenamefont
  {Balso}(2018)}]{sergeev2018horovod}%
  \BibitemOpen
  \bibfield  {author} {\bibinfo {author} {\bibfnamefont {Alexander}\
  \bibnamefont {Sergeev}}\ and\ \bibinfo {author} {\bibfnamefont {Mike~Del}\
  \bibnamefont {Balso}},\ }\bibfield  {title} {\enquote {\bibinfo {title}
  {Horovod: fast and easy distributed deep learning in {TensorFlow}},}\
  }\href@noop {} {\bibfield  {journal} {\bibinfo  {journal} {arXiv preprint
  arXiv:1802.05799}\ } (\bibinfo {year} {2018})}\BibitemShut {NoStop}%
\bibitem [{\citenamefont {Rasley}\ \emph {et~al.}(2020)\citenamefont {Rasley},
  \citenamefont {Rajbhandari}, \citenamefont {Ruwase},\ and\ \citenamefont
  {He}}]{deepSpeed}%
  \BibitemOpen
  \bibfield  {author} {\bibinfo {author} {\bibfnamefont {Jeff}\ \bibnamefont
  {Rasley}}, \bibinfo {author} {\bibfnamefont {Samyam}\ \bibnamefont
  {Rajbhandari}}, \bibinfo {author} {\bibfnamefont {Olatunji}\ \bibnamefont
  {Ruwase}}, \ and\ \bibinfo {author} {\bibfnamefont {Yuxiong}\ \bibnamefont
  {He}},\ }\enquote {\bibinfo {title} {Deepspeed: System optimizations enable
  training deep learning models with over 100 billion parameters},}\ in\ \href
  {https://doi.org/10.1145/3394486.3406703} {\emph {\bibinfo {booktitle}
  {Proceedings of the 26th ACM SIGKDD International Conference on Knowledge
  Discovery \& Data Mining}}}\ (\bibinfo  {publisher} {Association for
  Computing Machinery},\ \bibinfo {address} {New York, NY, USA},\ \bibinfo
  {year} {2020})\ p.\ \bibinfo {pages} {3505–3506}\BibitemShut {NoStop}%
\bibitem [{\citenamefont {Li}\ \emph {et~al.}(2020)\citenamefont {Li},
  \citenamefont {Zhao}, \citenamefont {Varma}, \citenamefont {Salpekar},
  \citenamefont {Noordhuis}, \citenamefont {Li}, \citenamefont {Paszke},
  \citenamefont {Smith}, \citenamefont {Vaughan}, \citenamefont {Damania},\
  and\ \citenamefont {Chintala}}]{li2020pytorch}%
  \BibitemOpen
  \bibfield  {author} {\bibinfo {author} {\bibfnamefont {Shen}\ \bibnamefont
  {Li}}, \bibinfo {author} {\bibfnamefont {Yanli}\ \bibnamefont {Zhao}},
  \bibinfo {author} {\bibfnamefont {Rohan}\ \bibnamefont {Varma}}, \bibinfo
  {author} {\bibfnamefont {Omkar}\ \bibnamefont {Salpekar}}, \bibinfo {author}
  {\bibfnamefont {Pieter}\ \bibnamefont {Noordhuis}}, \bibinfo {author}
  {\bibfnamefont {Teng}\ \bibnamefont {Li}}, \bibinfo {author} {\bibfnamefont
  {Adam}\ \bibnamefont {Paszke}}, \bibinfo {author} {\bibfnamefont {Jeff}\
  \bibnamefont {Smith}}, \bibinfo {author} {\bibfnamefont {Brian}\ \bibnamefont
  {Vaughan}}, \bibinfo {author} {\bibfnamefont {Pritam}\ \bibnamefont
  {Damania}}, \ and\ \bibinfo {author} {\bibfnamefont {Soumith}\ \bibnamefont
  {Chintala}},\ }\href@noop {} {\enquote {\bibinfo {title} {Pytorch
  distributed: Experiences on accelerating data parallel training},}\ }
  (\bibinfo {year} {2020}),\ \Eprint {http://arxiv.org/abs/2006.15704}
  {arXiv:2006.15704 [cs.DC]} \BibitemShut {NoStop}%
\bibitem [{\citenamefont {Laanait}\ \emph {et~al.}(2019)\citenamefont
  {Laanait}, \citenamefont {Romero}, \citenamefont {Yin}, \citenamefont
  {Young}, \citenamefont {Treichler}, \citenamefont {Starchenko}, \citenamefont
  {Borisevich}, \citenamefont {Sergeev},\ and\ \citenamefont
  {Matheson}}]{laanait2019exascale}%
  \BibitemOpen
  \bibfield  {author} {\bibinfo {author} {\bibfnamefont {Nouamane}\
  \bibnamefont {Laanait}}, \bibinfo {author} {\bibfnamefont {Joshua}\
  \bibnamefont {Romero}}, \bibinfo {author} {\bibfnamefont {Junqi}\
  \bibnamefont {Yin}}, \bibinfo {author} {\bibfnamefont {M~Todd}\ \bibnamefont
  {Young}}, \bibinfo {author} {\bibfnamefont {Sean}\ \bibnamefont {Treichler}},
  \bibinfo {author} {\bibfnamefont {Vitalii}\ \bibnamefont {Starchenko}},
  \bibinfo {author} {\bibfnamefont {Albina}\ \bibnamefont {Borisevich}},
  \bibinfo {author} {\bibfnamefont {Alex}\ \bibnamefont {Sergeev}}, \ and\
  \bibinfo {author} {\bibfnamefont {Michael}\ \bibnamefont {Matheson}},\
  }\bibfield  {title} {\enquote {\bibinfo {title} {Exascale deep learning for
  scientific inverse problems},}\ }\href@noop {} {\bibfield  {journal}
  {\bibinfo  {journal} {arXiv preprint arXiv:1909.11150}\ } (\bibinfo {year}
  {2019})}\BibitemShut {NoStop}%
\bibitem [{in_({\natexlab{b}})}]{in_prep_uncertainty_wp}%
  \BibitemOpen
  \enquote {\bibinfo {title} {{Uncertainty Quantification in Machine
  Learning}},}\ in\ \href@noop {} {\emph {\bibinfo {booktitle} {{Snowmass
  2022}}}},\ \bibinfo {note} {{in preparation}}\BibitemShut {NoStop}%
\bibitem [{\citenamefont {Wilkinson}\ \emph {et~al.}(2016)\citenamefont
  {Wilkinson}, \citenamefont {Dumontier}, \citenamefont {Aalbersberg},
  \citenamefont {Appleton}, \citenamefont {Axton}, \citenamefont {Baak},
  \citenamefont {Blomberg}, \citenamefont {Boiten}, \citenamefont
  {da~Silva~Santos}, \citenamefont {Bourne} \emph
  {et~al.}}]{wilkinson2016fair}%
  \BibitemOpen
  \bibfield  {author} {\bibinfo {author} {\bibfnamefont {Mark~D}\ \bibnamefont
  {Wilkinson}}, \bibinfo {author} {\bibfnamefont {Michel}\ \bibnamefont
  {Dumontier}}, \bibinfo {author} {\bibfnamefont {IJsbrand~Jan}\ \bibnamefont
  {Aalbersberg}}, \bibinfo {author} {\bibfnamefont {Gabrielle}\ \bibnamefont
  {Appleton}}, \bibinfo {author} {\bibfnamefont {Myles}\ \bibnamefont {Axton}},
  \bibinfo {author} {\bibfnamefont {Arie}\ \bibnamefont {Baak}}, \bibinfo
  {author} {\bibfnamefont {Niklas}\ \bibnamefont {Blomberg}}, \bibinfo {author}
  {\bibfnamefont {Jan-Willem}\ \bibnamefont {Boiten}}, \bibinfo {author}
  {\bibfnamefont {Luiz~Bonino}\ \bibnamefont {da~Silva~Santos}}, \bibinfo
  {author} {\bibfnamefont {Philip~E}\ \bibnamefont {Bourne}},  \emph {et~al.},\
  }\bibfield  {title} {\enquote {\bibinfo {title} {The fair guiding principles
  for scientific data management and stewardship},}\ }\href@noop {} {\bibfield
  {journal} {\bibinfo  {journal} {Scientific data}\ }\textbf {\bibinfo {volume}
  {3}},\ \bibinfo {pages} {1--9} (\bibinfo {year} {2016})}\BibitemShut
  {NoStop}%
\bibitem [{in_({\natexlab{c}})}]{in_prep_education_wp}%
  \BibitemOpen
  \enquote {\bibinfo {title} {{Data Science \& Machine Learning Education in
  HEP}},}\ in\ \href@noop {} {\emph {\bibinfo {booktitle} {{Snowmass 2022}}}},\
  \bibinfo {note} {{in preparation}}\BibitemShut {NoStop}%
\bibitem [{\citenamefont {Feickert}\ and\ \citenamefont
  {Nachman}(2021)}]{Feickert:2021ajf}%
  \BibitemOpen
  \bibfield  {author} {\bibinfo {author} {\bibfnamefont {Matthew}\ \bibnamefont
  {Feickert}}\ and\ \bibinfo {author} {\bibfnamefont {Benjamin}\ \bibnamefont
  {Nachman}},\ }\bibfield  {title} {\enquote {\bibinfo {title} {{A Living
  Review of Machine Learning for Particle Physics}},}\ }\href@noop {} {\
  (\bibinfo {year} {2021})},\ \Eprint {http://arxiv.org/abs/2102.02770}
  {arXiv:2102.02770 [hep-ph]} \BibitemShut {NoStop}%
\bibitem [{\citenamefont {Boehnlein}\ \emph {et~al.}(2021)\citenamefont
  {Boehnlein} \emph {et~al.}}]{Boehnlein:2021eym}%
  \BibitemOpen
  \bibfield  {author} {\bibinfo {author} {\bibfnamefont {Amber}\ \bibnamefont
  {Boehnlein}} \emph {et~al.},\ }\bibfield  {title} {\enquote {\bibinfo {title}
  {{Artificial Intelligence and Machine Learning in Nuclear Physics}},}\
  }\href@noop {} {\  (\bibinfo {year} {2021})},\ \Eprint
  {http://arxiv.org/abs/2112.02309} {arXiv:2112.02309 [nucl-th]} \BibitemShut
  {NoStop}%
\bibitem [{\citenamefont {Tanaka}\ \emph {et~al.}(2021)\citenamefont {Tanaka},
  \citenamefont {Tomiya},\ and\ \citenamefont {Hashimoto}}]{tanaka2021deep}%
  \BibitemOpen
  \bibfield  {author} {\bibinfo {author} {\bibfnamefont {Akinori}\ \bibnamefont
  {Tanaka}}, \bibinfo {author} {\bibfnamefont {Akio}\ \bibnamefont {Tomiya}}, \
  and\ \bibinfo {author} {\bibfnamefont {K{\=o}ji}\ \bibnamefont {Hashimoto}},\
  }\href@noop {} {\emph {\bibinfo {title} {Deep Learning and Physics}}}\
  (\bibinfo  {publisher} {Springer},\ \bibinfo {year} {2021})\BibitemShut
  {NoStop}%
\bibitem [{\citenamefont {Cuoco}\ \emph {et~al.}(2021)\citenamefont {Cuoco}
  \emph {et~al.}}]{Cuoco:2020ogp}%
  \BibitemOpen
  \bibfield  {author} {\bibinfo {author} {\bibfnamefont {Elena}\ \bibnamefont
  {Cuoco}} \emph {et~al.},\ }\bibfield  {title} {\enquote {\bibinfo {title}
  {{Enhancing Gravitational-Wave Science with Machine Learning}},}\ }\href
  {\doibase 10.1088/2632-2153/abb93a} {\bibfield  {journal} {\bibinfo
  {journal} {Mach. Learn. Sci. Tech.}\ }\textbf {\bibinfo {volume} {2}},\
  \bibinfo {pages} {011002} (\bibinfo {year} {2021})},\ \Eprint
  {http://arxiv.org/abs/2005.03745} {arXiv:2005.03745 [astro-ph.HE]}
  \BibitemShut {NoStop}%
\bibitem [{\citenamefont {Baron}(2019)}]{baron2019machine}%
  \BibitemOpen
  \bibfield  {author} {\bibinfo {author} {\bibfnamefont {Dalya}\ \bibnamefont
  {Baron}},\ }\href@noop {} {\enquote {\bibinfo {title} {Machine learning in
  astronomy: a practical overview},}\ } (\bibinfo {year} {2019}),\ \Eprint
  {http://arxiv.org/abs/1904.07248} {arXiv:1904.07248 [astro-ph.IM]}
  \BibitemShut {NoStop}%
\bibitem [{\citenamefont {Bedolla-Montiel}\ \emph {et~al.}(2021)\citenamefont
  {Bedolla-Montiel}, \citenamefont {Padierna},\ and\ \citenamefont {Casta\~neda
  Priego}}]{Bedolla-Montiel:2020rio}%
  \BibitemOpen
  \bibfield  {author} {\bibinfo {author} {\bibfnamefont {Edwin~A.}\
  \bibnamefont {Bedolla-Montiel}}, \bibinfo {author} {\bibfnamefont
  {Luis~Carlos}\ \bibnamefont {Padierna}}, \ and\ \bibinfo {author}
  {\bibfnamefont {Ram\'on}\ \bibnamefont {Casta\~neda Priego}},\ }\bibfield
  {title} {\enquote {\bibinfo {title} {{Machine Learning for Condensed Matter
  Physics}},}\ }\href {\doibase 10.1088/1361-648X/abb895} {\bibfield  {journal}
  {\bibinfo  {journal} {J. Phys. Condens. Matter}\ }\textbf {\bibinfo {volume}
  {33}},\ \bibinfo {pages} {053001} (\bibinfo {year} {2021})},\ \Eprint
  {http://arxiv.org/abs/2005.14228} {arXiv:2005.14228 [physics.comp-ph]}
  \BibitemShut {NoStop}%
\bibitem [{\citenamefont {Brunton}\ \emph {et~al.}(2020)\citenamefont
  {Brunton}, \citenamefont {Noack},\ and\ \citenamefont
  {Koumoutsakos}}]{doi:10.1146/annurev-fluid-010719-060214}%
  \BibitemOpen
  \bibfield  {author} {\bibinfo {author} {\bibfnamefont {Steven~L.}\
  \bibnamefont {Brunton}}, \bibinfo {author} {\bibfnamefont {Bernd~R.}\
  \bibnamefont {Noack}}, \ and\ \bibinfo {author} {\bibfnamefont {Petros}\
  \bibnamefont {Koumoutsakos}},\ }\bibfield  {title} {\enquote {\bibinfo
  {title} {Machine learning for fluid mechanics},}\ }\href {\doibase
  10.1146/annurev-fluid-010719-060214} {\bibfield  {journal} {\bibinfo
  {journal} {Annual Review of Fluid Mechanics}\ }\textbf {\bibinfo {volume}
  {52}},\ \bibinfo {pages} {477--508} (\bibinfo {year} {2020})},\ \Eprint
  {http://arxiv.org/abs/https://doi.org/10.1146/annurev-fluid-010719-060214}
  {https://doi.org/10.1146/annurev-fluid-010719-060214} \BibitemShut {NoStop}%
\bibitem [{\citenamefont {Kochkov}\ \emph {et~al.}(2021)\citenamefont
  {Kochkov}, \citenamefont {Smith}, \citenamefont {Alieva}, \citenamefont
  {Wang}, \citenamefont {Brenner},\ and\ \citenamefont
  {Hoyer}}]{Kochkove2101784118}%
  \BibitemOpen
  \bibfield  {author} {\bibinfo {author} {\bibfnamefont {Dmitrii}\ \bibnamefont
  {Kochkov}}, \bibinfo {author} {\bibfnamefont {Jamie~A.}\ \bibnamefont
  {Smith}}, \bibinfo {author} {\bibfnamefont {Ayya}\ \bibnamefont {Alieva}},
  \bibinfo {author} {\bibfnamefont {Qing}\ \bibnamefont {Wang}}, \bibinfo
  {author} {\bibfnamefont {Michael~P.}\ \bibnamefont {Brenner}}, \ and\
  \bibinfo {author} {\bibfnamefont {Stephan}\ \bibnamefont {Hoyer}},\
  }\bibfield  {title} {\enquote {\bibinfo {title} {Machine
  learning{\textendash}accelerated computational fluid dynamics},}\ }\href
  {\doibase 10.1073/pnas.2101784118} {\bibfield  {journal} {\bibinfo  {journal}
  {Proceedings of the National Academy of Sciences}\ }\textbf {\bibinfo
  {volume} {118}} (\bibinfo {year} {2021}),\ 10.1073/pnas.2101784118},\ \Eprint
  {http://arxiv.org/abs/https://www.pnas.org/content/118/21/e2101784118.full.pdf}
  {https://www.pnas.org/content/118/21/e2101784118.full.pdf} \BibitemShut
  {NoStop}%
\bibitem [{\citenamefont {Sch{\"u}tt}\ \emph {et~al.}(2020)\citenamefont
  {Sch{\"u}tt}, \citenamefont {Chmiela}, \citenamefont {von Lilienfeld},
  \citenamefont {Tkatchenko}, \citenamefont {Tsuda},\ and\ \citenamefont
  {M{\"u}ller}}]{schutt2020machine}%
  \BibitemOpen
  \bibfield  {author} {\bibinfo {author} {\bibfnamefont {Kristof~T}\
  \bibnamefont {Sch{\"u}tt}}, \bibinfo {author} {\bibfnamefont {Stefan}\
  \bibnamefont {Chmiela}}, \bibinfo {author} {\bibfnamefont {O~Anatole}\
  \bibnamefont {von Lilienfeld}}, \bibinfo {author} {\bibfnamefont {Alexandre}\
  \bibnamefont {Tkatchenko}}, \bibinfo {author} {\bibfnamefont {Koji}\
  \bibnamefont {Tsuda}}, \ and\ \bibinfo {author} {\bibfnamefont
  {Klaus-Robert}\ \bibnamefont {M{\"u}ller}},\ }\bibfield  {title} {\enquote
  {\bibinfo {title} {Machine learning meets quantum physics},}\ }\href@noop {}
  {\bibfield  {journal} {\bibinfo  {journal} {Lecture Notes in Physics}\ }
  (\bibinfo {year} {2020})}\BibitemShut {NoStop}%
\end{thebibliography}%

\end{document}